\documentclass[11pt,a4paper,reqno,intlimits,sumlimits]{amsart}

\usepackage{amssymb,amsmath}
\usepackage{color}
\usepackage[mathscr]{euscript}
\usepackage{xspace}
\usepackage{epsfig,rotating}

\begin{document}

\frenchspacing

\theoremstyle{plain}
\newtheorem{theorem}{Theorem}[section]
\newtheorem{lemma}[theorem]{Lemma}
\newtheorem{proposition}[theorem]{Proposition}
\newtheorem{corollary}[theorem]{Corollary}

\theoremstyle{definition}
\newtheorem{remark}[theorem]{Remark}
\newtheorem{definition}[theorem]{Definition}
\newtheorem{assumption}{Assumption}
\newtheorem{example}[theorem]{Example}
\renewcommand{\theequation}{\thesection.\arabic{equation}}
\numberwithin{equation}{section}

\renewcommand{\thetable}{\thesection.\arabic{table}}
\numberwithin{table}{section}

\renewcommand{\thefigure}{\thesection.\arabic{figure}}
\numberwithin{figure}{section}

\def\stackrelboth#1#2#3{\mathrel{\mathop{#2}\limits^{#1}_{#3}}}
\def\scal#1#2{\langle #1,#2 \rangle}
\def\P{\ensuremath{\mathrm{I\kern-.2em P}}}
\def\E{\mathrm{I\kern-.2em E}}

\def\bF{\mathbf{F}}
\def\F{\ensuremath{\mathcal{F}}}
\def\R{\ensuremath{\mathbb{R}}}
\def\C{\ensuremath{\mathbb{C}}}
\def\bt{\ensuremath{\mathbf{T}}}

\def\Rmz{\R\setminus\{0\}}
\def\Rdmz{\R^d\setminus\{0\}}
\def\Rnmz{\R^n\setminus\{0\}}

\def\Rp{\mathbb{R}_{\geqslant0}}

\def\lev{L\'{e}vy\xspace}
\def\lib{LIBOR\xspace}
\def\lk{L\'{e}vy--Khintchine\xspace}
\def\smmg{semimartingale\xspace}
\def\smmgs{semimartingales\xspace}
\def\mg{martingale\xspace}
\def\tih{time-inhomoge\-neous\xspace}

\def\eqlaw{\ensuremath{\stackrel{\mathrrefersm{d}}{=}}}

\def\ud{\ensuremath{\mathrm{d}}}
\def\dt{\ud t}
\def\ds{\ud s}
\def\dx{\ud x}
\def\dy{\ud y}
\def\dsdx{\ensuremath{(\ud s, \ud x)}}
\def\dtdx{\ensuremath{(\ud t, \ud x)}}

\def\intrr{\ensuremath{\int_{\R}}}

\def\ott{{0\leq t\leq T_*}}

\def\e{\mathrm{e}}
\def\eps{\epsilon}

\def\t0i{_i(0)}
\def\tsi{_i(s)}
\def\tti{_i(t)}

\def\ykl{\ensuremath{Y_{kl}}\xspace}
\def\kl{{kl}}
\def\oV{\overline{V}}
\def\mt{\widetilde{\mu}}
\def\half{\frac{1}{2}}

\title[Approximations to L\'evy LIBOR models]{Efficient and accurate log-L\'evy
       approximations to L\'evy driven \lib models}

\author[A. Papapantoleon]{Antonis Papapantoleon}
\author[J. Schoenmakers]{John Schoenmakers}
\author[D. Skovmand]{David Skovmand}

\address{Institute of Mathematics, TU Berlin, Stra\ss e des 17. Juni 136,
         10623 Berlin, Germany}
\email{papapan@math.tu-berlin.de}

\address{Weierstrass Institute for Applied Analysis and Stochastics, Mohrenstr.
         39, 10117 Berlin, Germany}
\email{schoenma@wias-berlin.de}

\address{Department of Economics and Business, Aarhus University, Fuglesangs
All\'e 4,
         8210 Aarhus V, Denmark}
\email{davids@asb.dk}
\date{}
\thanks{J. S. acknowledges the financial support by the DFG Research Center
        \textsc{Matheon} ``Mathematics for Key Technologies'' in Berlin.}
\date{}\maketitle
\pagestyle{myheadings}\frenchspacing

\begin{abstract}
The \lib market model is very popular for pricing interest rate derivatives, but
is known to have several pitfalls. In addition, if the model is driven by a jump
process, then the complexity of the drift term is growing exponentially fast (as
a function of the tenor length). In this work, we consider a \lev-driven LIBOR
model and aim at developing accurate and efficient log-\lev approximations for
the dynamics of the rates. The approximations are based on truncation of the
drift term and Picard approximation of suitable processes. Numerical experiments
for FRAs, caps, swaptions and sticky ratchet caps show that the approximations
perform very well. In addition, we also consider the log-\lev approximation of
annuities, which offers good approximations for high volatility regimes.
\end{abstract}

\section{Introduction}
\label{intro}

The \lib market model (LMM) has become a standard model for the pricing of
interest rate derivatives in recent years, because the evolution of discretely
compounded, market-observable forward rates is modeled directly and not deduced
from the evolution of unobservable factors, as is the case in short rate and
forward rate (HJM) models. See \cite{MiltersenSandmannSondermann97},
\cite{BraceGatarekMusiela97} and \cite{Jamshidian97} for the seminal
papers in \lib modeling. In addition, the lognormal \lib model provides a
theoretical justification to the market practice of pricing caps according to
Black's formula (cf. \cite{Black76}). However, despite its apparent
popularity, the \lib market model has certain well-known pitfalls.

An interest rate model is typically calibrated to the implied volatility surface
from the cap market and the correlation structure of at-the-money swaptions. The
implied volatility from caplets has a ``smile'' shape as a function of strike,
while its term structure is typically decreasing. The standard lognormal LMM
cannot be calibrated adequately to the observed market data. Therefore, several
extensions of the LMM have been proposed in the literature using
jump-diffusions, \lev processes or general semimartingales as the driving motion
(cf. e.g. \cite{GlassermanKou03}, \cite{EberleinOezkan05},
\cite{Jamshidian99}), or incorporating stochastic volatility effects (cf. e.g.
\cite{AndersenBrothertonRatcliffe05}, \cite{Wu_Zhang_2006},
\cite{BelomestnyMathewSchoenmakers06}).

The dynamics of \lib models are typically not tractable under different forward
measures, due to the random terms that enter the dynamics of LIBOR rates. In
particular, if the driving process is a diffusion process or a general
semimartingale, then the dynamics of \lib rates are not tractable even under
their own forward measures. Consequently, even caplets cannot be priced exactly
in ``closed form'' (meaning, e.g. by Fourier methods), let alone swaptions and
other multi-LIBOR products. In order to calibrate the model, closed form
solutions are necessary, and these are typically involving approximations.

The standard approximation is the so-called ``frozen drift'' approximation; it
was first proposed by \cite{BraceGatarekMusiela97} for the pricing of
swaptions and has been used by several authors ever since. The frozen drift
approximation typically leads to closed-form solutions for caplet pricing in
realistic LIBOR models, see \cite{EberleinOezkan05} and
\cite{BelomestnyMathewSchoenmakers06}. Although some authors
(\cite{BraceDunBarton01}, \cite{DunBartonSchloegl01} and
\cite{Schloegl02}) argue that freezing the drift is justified in the
lognormal LMM, it is shown that it does not yield acceptable results for
exotic derivatives and longer time horizons, see e.g.
\cite{KurbanmuradovSabelfeldSchoenmakers}. Therefore, several alternative
approximations have been developed in the literature. In one line of research,
\cite{KurbanmuradovSabelfeldSchoenmakers} and \cite{DanilukGatarek05}
have derived lognormal approximations to the forward LIBOR dynamics (for
deterministic volatility structures). Other authors have been using linear
interpolations and predictor-corrector Monte Carlo methods to get a more
accurate discretization of the drift term (cf. e.g.
\cite{HunterJaeckelJoshi01} and \cite{GlassermanZhao00}). We refer the
reader to \cite{JoshiStacey08} and \cite[Ch. 10]{GatarekBachertMaksymiuk06}
for a detailed overview of that literature, some new approximation schemes and
numerical experiments. Although most of this literature focuses on the lognormal
LMM, \cite{GlassermanMerener03} and
\cite{GlassermanMerener03b}) have developed approximation schemes for the
pricing of caps and swaptions in jump-diffusion \lib market models, based on
freezing the drift. 

In this article, we consider a \lib market model driven by a \lev process and
aim at deriving efficient and more accurate log-\lev approximations (compared to
the ``frozen drift'' approximation, for instance). As a main result, we develop
log-\lev LIBOR approximations which may be represented as a {\em deterministic}
drift term plus a stochastic integral of a deterministic function with respect
to a \lev process. In particular, in the context of Monte Carlo simulation the
drift term can be computed outside the Monte Carlo loop, while the stochastic
integrals can be computed efficiently for each trajectory. In contrast, standard
Euler stepping of the original LIBOR SDE involves, for each LIBOR trajectory, an
accurate computation of a complex-structured random drift term at each Euler
step and is therefore significantly more time-consuming\footnote{In a previous
unpublished manuscript by the first and third author
\cite{PapapantoleonSkovmand10} the efficiency of the standard Euler approach was
improved to some extend also, but there was still a costly random drift
involved.}. Theoretical investigations as well as numerical experiments
show that the log-\lev approximations are both fast and accurate when the LIBOR
volatilities are not too high, and thus provide an effective alternative to
simulation methods based on standard Euler discretizations. Finally, as a
generalization of \cite{GatarekBachertMaksymiuk06}, we derive log-\lev
approximations for annuity terms, which allow for pricing options in high
volatility regimes.

The article is structured as follows: in section \ref{model} we review the
\lev-driven \lib model, in section \ref{LogLevy} we construct the log-\lev
approximations to the model and in section \ref{error} we provide some error
estimates. Section \ref{numerics} demonstrates numerically the effect of the
approximations, while section \ref{annuity-section} deals with an approximation
of annuities. The final section provides some recommendations on the
construction of multi-dimensional \lev \lib models, while the appendices collect
various calculations.

\section{L\'evy \lib framework}
\label{model}

Let $0=T_0<T_{1}<\cdots<T_{N}<T_{N+1}=T_*$ denote a discrete tenor structure
where $\delta_i=T_{i+1}-T_i$, $i=0,1,\dots,N,$ are the so called day-count
fractions. For this tenor structure we consider an arbitrage free system of zero
coupon bond processes $B_i,$ $i=1,\ldots,N+1,$ on a filtered probability space
$(\Omega,\F,(\F_t)_{0\le t\le T_*},\P_*),$ where $\P_*:=\P_{N+1}$ is a numeraire
measure connected with the terminal bond $B_{N+1}$. From this bond system we may
deduce a forward rate system, also called \lib rate system, defined by
\begin{equation}\label{Ld}
L_{i}(t):=\frac{1}{\delta_{i}}\left(\frac{B_{i}(t)}{B_{i+1}(t)}-1\right),
 \quad 0\leq t\le T_{i},\; 1\leq i\leq N.
\end{equation}
\(L_{i}\) is the annualized effective forward rate contracted at date \(t\le
T_i\) for the period \([T_{i},T_{i+1}]\). \cite{Jamshidian99} derived a general
representation for the \lib dynamics in a semimartingale framework. In this
article we consider a \lev \lib framework as constructed by
\cite{EberleinOezkan05}; see also \cite{GlassermanKou03} and
\cite{BelomestnySchoenmakers06} for jump-diffusion settings.

Consider a standard Brownian motion $W$ in $\mathbb{R}^m$, $m\le N$, a bounded
deterministic nonnegative scalar function $\alpha(s)$, $s\in[0,T_*]$, and a
random measure $\mu$ on $[0,T_*]\times\mathbb{R}^{m}$ with
$\P_{\ast}$-compensator $F(s,\dx)\ds$, where $\mu$ and $W$ are mutually
independent. Let $H=(H(t))_{\ott}$ be a \tih \lev process  with canonical
decomposition
\begin{equation}\label{h*}
H(t)
 = \int_{0}^{t}\sqrt{\alpha(s)}\ud W(s)
 +\int_{0}^{t}\int_{\mathbb{R}^{m}}x(\mu\dsdx-F(s,\dx)\ds).
\end{equation}
We denote by $\mt$ the compensated random measure of the jumps of $H$, that is
$\mt\dsdx:=\mu\dsdx-F(s,\dx)\ds$. In order to avoid truncation conventions we
assume that $F$ satisfies the (stronger than usual) integrability condition
\begin{align*}
\int_{0}^{T_{\ast}}\int_{\mathbb{R}^{m}}
 \left( \left\Vert x\right\Vert \wedge\left\Vert x\right\Vert^2 \right)
   F(s,\dx)\ds<\infty.
\end{align*}
We further assume that
\begin{align}\label{EMcondition}
\int_{0}^{T_{\ast}}\int_{\|x\|>1}
  \exp\big( u^\mathsf{T}x\big) F(s,\dx)\ds<\infty,
\end{align}
for all $\|u\|\le(1+\varepsilon)\overline{M}$, with $\overline{M},\varepsilon>0$
constants. Thus, by construction, the process $(H(t))_\ott$ is a
$\P_*$-martingale. The cumulant generating function of $H(t)$, $t\in[0,T_*]$, is
provided by
\begin{align}\label{cumulant-H0}
\ln \E\big[\e^{u^\mathsf{T}H(t)}\big]
 = \int_0^t \kappa_s(u)\ds,
\end{align}
where
\begin{align}\label{cumulant-H}
\kappa_s(u)
 = \frac{\alpha(s)}{2}\|u\|^2
 + \int_{\R^m}
    \big(\e^{u^{\mathsf{T}}x}-1-u^{\mathsf{T}}x\big)F(s,\dx).
\end{align}

Along with the \lev martingale \eqref{h*} we introduce a set of bounded
deterministic vector-valued functions $\lambda_{i}(s)\in\mathbb{R}^{m}$,
$i=1,\dots,N,$ usually called \emph{loading factors}. In order to avoid local
redundances we assume that the matrix $[\lambda_{1},\ldots,\lambda_{N}](s)$ has
full rank $m$ for all $s\in[0,T_*]$. Moreover, we assume that
$\|\lambda_i(s)\|\le~\overline{M}$, for all $i$, and $\|\sum_i\lambda_i(s)\|\le
\overline{M}$, for all $s\in[0,T_*]$.

The \lev martingale and the set of loading factors then constitute an
arbitrage free \lib system consistent with \eqref{Ld}, whose dynamics under the
terminal measure $\P_{*}$ are given by
\begin{equation}\label{logl}
L_{i}(t)
 = L_{i}(0)\exp\left( \int_{0}^{t}b_{i}(s)\ds
    +\int_{0}^{t}\lambda_{i}^{\mathsf{T}}(s) \ud H(s)\right),
\end{equation}
$i=1,\dots,N$, where the drift terms in the exponent are given by
\begin{align}\label{dr}
b_{i}
 &=-\frac{1}{2}\alpha\left\vert \lambda_{i} \right\vert^{2}
   -\sum_{j=i+1}^{N}\frac{\delta_{j}L_{j-}}{1+\delta_{j}L_{j-}}
     \alpha\lambda_{i}^{\mathsf{T}}\lambda_{j}\\
 &\quad  \nonumber
   -\int_{\mathbb{R}^{m}}\left( \left(\e^{\lambda_{i}^{\mathsf{T}}x}-1\right)
     \prod_{j=i+1}^{N}
  \left(1 + \frac{\delta_{j}L_{j-}\left(\e^{\lambda_{i}^{\mathsf{T}}x}-1\right)}
      {1+\delta_{j}L_{j-}}\right) -\lambda_{i}^{\mathsf{T}}x\right)F(\cdot,\dx);
\end{align}
for details see \cite{EberleinOezkan05}. For notational convenience, we set
$L_{j-}(s):=L_{j}(s-)$ in \eqref{dr}, while the time variable is suppressed.

Due to the drift term \eqref{dr}, a straightforward Monte Carlo simulation of
\eqref{logl} would involve a numerical integration at each time step, since the
random terms $\frac{\delta_{j}L_{j-}}{1+\delta_{j}L_{j-}}$ appear under the
integral sign. In order to overcome this problem, we will re-express the drift
in terms of random quotients multiplied with cumulants of the driving process.
We have that
\begin{align}\label{drift-effi}
b_{i}
 &=-\kappa(\lambda_{i})
   -\sum_{j=i+1}^{N}\frac{\delta_{j}L_{j-}}{1+\delta_{j}L_{j-}}
     \alpha\lambda_{i}^{\mathsf{T}}\lambda_{j} \nonumber \\ \nonumber
 &\quad
   - \sum_{p=1}^{N-i}\sum_{i<j_{1}<\cdots<j_{p}\leq N}
       \frac{\delta_{j_{1}}L_{j_{1}-}}{1+\delta_{j_{1}}L_{j_{1}-}}
       \cdots \frac{\delta_{j_{p}}L_{j_{p}-}}{1+\delta_{j_{p}}L_{j_{p}-}}\\
 &\qquad
    \times \sum_{q=1}^{p+1}(-1)^{p+q+1}
       \sum_{0\leq r_{1}<\cdots<r_{q}\leq p}
        \widehat\kappa(\lambda_{j_{r_1}}+\cdots+\lambda_{j_{r_1}});
\end{align}
the derivation is deferred to Appendix \ref{app-drift}, for brevity. Here
$\widehat\kappa$ denotes the part of the cumulant $\kappa$ stemming from the
jumps of $L$, that is
\begin{align}\label{cumulant-H-kumps}
\widehat\kappa_s(u)
 = \int_{\R^m}
    \big(\e^{u^{\mathsf{T}}x}-1-u^{\mathsf{T}}x\big)F(s,\dx).
\end{align}
Therefore, we can now avoid the numerical integration when simulating \lib
rates. However, another problem becomes apparent in this representation: the
number of terms to be computed in \eqref{drift-effi} grows exponentially fast as
a function of the number of \lib rates $N$, namely it has order $O(2^N)$.

\begin{remark}
In a practically applicable model, the loading factors $\lambda_i$ may be
decomposed as follows:
\begin{gather*}
\lambda_i(t)=c_i g(T_i-t)e_{i-m(t)}\in \mathbb{R}^m,\\
 m(t):=\inf\{i: T_i\ge t\},
 \quad \|e_i\|=1,\quad  e^{\mathsf{T}}_i e_j=\rho_{ij},
 \quad 1\le i,j \le N,
\end{gather*}
for constants $c_i>0$, some (e.g. parametric) scalar function $g>0$, and a
correlation structure $(\rho_{ij})$ which resembles the correlations between
forward LIBORs observed in the market. For instance, $(\rho_{ij})$ may be
obtained as a rank-$m$ approximation of a suitably parameterized full rank-$N$
correlation structure; see \cite{Schoenmakers05} for details. Further, the
scalar function $\alpha$ may be taken as a constant that controls the influence
of the Wiener noise with respect to the jump noise.
\end{remark}

\begin{remark}
Using semi-analytic pricing methods based on Fourier transforms, the
L\'evy-driven \lib model may be calibrated to caplet volatilities for different
strikes and maturities in the spirit of \cite{BelomestnySchoenmakers06},
\cite{EberleinKluge06} and \cite{Beinhofer_Eberlein_Janssen_Polley_2011}.
\end{remark}

\begin{remark}\label{remP}
The \lev-driven \lib model is constructed under the terminal measure $\P_{N+1}$
in this paper, for definiteness. As an alternative, for products with shorter
maturity for instance, one may consider for some $T_{\widetilde{N}}<T_{N+1}$, a
\lev-driven \lib model for $t\leq T_{\widetilde{N}}$ under the measure
$\P_{\widetilde{N}}$, with respect to the numeraire bond $B_{\widetilde{N}}$.
Another possibility is to consider as numeraire the spot \lib rolling over
account
\begin{align*}
B_{\circ}(0) &:= 1, \text{ \ \ }
B_{\circ}(t):=\frac{B_{m(t)}(t)}{B_{1}(0)} 
 \prod\limits_{i=1}^{m(t)-1}(1+\delta_{i}L_{i}(T_{i})),\\
m(t)  & :=\min\{m:T_{m}\geq t\}, \text{ \ \ } 0<t\leq T_{N+1},
\end{align*}
and the numeraire measure $\P_{\circ}$ associated with it. If one prefers to
work in one of these other measures, the drift term \eqref{dr} has to be
modified in the following way: for the Libor model in the measure
$\P_{\widetilde{N}},$  replace in \eqref{dr}, if $i\leq\widetilde{N},$ the sum
$-\sum_{j=i+1}^{N}$ and the product ${\prod_{j=i+1}^{N}}$ by
$-\sum_{j=i+1}^{\widetilde{N}-1}$ and $\prod_{j=i+1}^{\widetilde{N}-1}$
respectively, and if $i>\widetilde{N}$ $,$ by $\sum_{j=\widetilde{N}}^{i}$
and $1/{\prod_{j=\widetilde{N}}^{i}}$ respectively. Likewise, for a \lib model
in the measure $\P_{\circ},$ replace in \eqref{dr} $-\sum_{j=i+1}^{N}$ by $\
\sum_{j=m(t)}^{i}$ and the product ${\prod_{j=i+1}^{N}}$ by
$1/{\prod_{j=m(t)}^{i}}$. We refer to Jamshidian (1999) for more details. The
proper choice of a numeraire measure under which the \lev-driven \lib model
is constructed may depend on the set of LIBORs involved in a particular
(structured) product which has to be evaluated by simulation. In principle, one
should choose the measure in such a way that the respective sum and product in
the drift \eqref{dr} involve as few terms as possible. 
\end{remark}

\section{Efficient and accurate log-\lev approximations}
\label{LogLevy}

The aim of this section is to derive efficient and accurate log-\lev
approximations for the dynamics of the \lib rates under the terminal measure.
This is based on an appropriate approximation of the drift term, cf. \eqref{dr},
which has two pillars:
\begin{enumerate}
\item expansion and truncation of the drift term,
\item Picard approximation of suitably defined processes.
\end{enumerate}
We will first provide an overview of the approximation argument, and then
present the full details in some particular cases.

\subsection{Outline of the method}

Let us denote the log-LIBOR rates by $G_i$. They are defined via
\[
G_i(t):= \log L_i(t),
\]
and satisfy the integrated linear SDE, see \eqref{logl},
\begin{equation}\label{intL}
G_i(t) = G_i(0) + \int_0^t b_i(s) \ds + \int_0^t \lambda_i^{\mathsf{T}}(s)\ud
H(s),
\end{equation}
$0\le t\le T_i$, $1\le i\le N$. The semimartingale characteristics of $G_i$ are
\begin{align}\label{G-triplet}
B^i &= \int\nolimits_0^\cdot b_i(s)\ud s \nonumber\\
C^i &= \int\nolimits_0^\cdot |\lambda_i|^2(s)\alpha(s)\ud s \\
\int\nolimits_0^\cdot\int\nolimits_{\R} 1_A(x)F^i(s,\dx)\ds
  &= \int\nolimits_0^\cdot\int\nolimits_{\R^m}
     1_A\big(\lambda_i^{\mathsf{T}}(s)x\big)F(s,\dx)\ds, \nonumber
\end{align}
where $A\in\mathcal{B}(\R\setminus\{0\})$.

Inspired by the lognormal approximation developed by
\cite{KurbanmuradovSabelfeldSchoenmakers} in the context of the lognormal
\lib market model, we will derive log-\lev approximations for the dynamics of
$L_i$, or equivalently \lev approximations for the dynamics of $G_i$. The
standard remedy for the numerical problems arising in LMMs is to ``freeze the
drift'', that is to replace the random terms in \eqref{dr} -- or
\eqref{drift-effi} -- by their deterministic initial values. In the present
model, this obviously leads to a log-\lev approximation, which however is not
accurate enough.

The method for deriving \textit{efficient} and \textit{accurate} log-\lev
approximations we propose can be summarized in the following steps:
\begin{itemize}
\item consider the different product terms
      $\frac{\delta_{j_{1}}L_{j_{1}}}{1+\delta_{j_{1}}L_{j_{1}}}\cdots
       \frac{\delta_{j_{p}}L_{j_{p}}}{1+\delta_{j_{p}}L_{j_{p}}}
       =:X_{j_1\dots j_p}$
      in \eqref{drift-effi}, where $i+1\le j_1<\dots<j_p\le N$;
\item define functions $h:\R^{j_p}\to\R$ such that
      $$h(G_{j_1},\dots,G_{j_p})=X_{j_1\dots j_p};$$
\item apply It\^o's formula to $X_{j_1\dots j_p}$, which leads to an SDE of the
      form
      \begin{align}
      \ud X_{j_1\dots j_p}(s) &= A_{j_1\dots j_p}(s,L(s))\ds
        + B_{j_1\dots j_p}(s,L(s))^{\mathsf{T}}\ud W(s) \notag\\
       &\quad+ \int_{\R^m} C_{j_1\dots j_p}(s,x,L(s))\mt\dsdx,
      \end{align}
      with $L=[L_1,\dots,L_N]$;
\item use the first step of a Picard iteration to approximate $X_{j_1\dots j_p}$
      by the \lev process
      \begin{align}\label{OM-levy}
       &X_{j_1\dots j_p}^{(1)}(t) = X_{j_1\dots j_p}(0)
       + \int_0^t A_{j_1\dots j_p}(s,L(0))\ds \\ \notag &
       + \int_0^t B_{j_1\dots j_p}(s,L(0))^{\mathsf{T}}\ud W_s
       + \int_0^t\int_{\R^m} C_{j_1\dots j_p}(s,x,L(0))\mt\dsdx;
      \end{align}
\item plug the \lev processes $X_{j_1\dots j_p}^{(1)}$ into $b_i$, cf.
      \eqref{drift-effi}, which leads to a \lev approximation for $b_i$;
\item finally, integrate by parts to deduce a \lev approximation for $G_i$
      of the form
\begin{align*}
G_i(t)
 &\approx \widehat{G}_i(0,t) + \int_{0}^{t} \mathrm{H}(t,s)\ds
  + \int_{0}^{t}\Theta^{\mathsf{T}}(t,s)\ud W(s)
  + \int_{0}^{t} \mathrm{I}(t,s,x)\mt\dsdx,
\end{align*}
where $\mathrm{H}, \Theta$ and $\mathrm{I}$ are deterministic, time-dependent
functions.
\end{itemize}
The main advantage of the above approximations is that they can be simulated
efficiently, as explained in section \ref{eff-sim}. Moreover, their
characteristic functions can be given in closed form.

\begin{remark}
Note that the ``frozen drift'' approximation can be easily embedded in this
scheme. It corresponds to using just the initial values $X_{j_1\dots j_p}(0)$
instead of the \lev process $X_{j_1\dots j_p}^{(1)}$ in \eqref{OM-levy}.
\end{remark}

\subsection{Log-\lev approximation schemes}
\label{effi}

In the sequel, we are going to follow this recipe for deriving efficient and
accurate log-\lev approximations, and present the full details of the method.
However, we will first truncate the drift terms at the second order, in order to
reduce the number of terms that need to be calculated.

\textbf{1.}
The first step is to expand and truncate the drift term at the second order;
these computations have been deferred to Appendix \ref{app-drift} for brevity,
see \eqref{so_int}. We will approximate $b_i$ by $b_i''$, where
\begin{align}\label{drift-2}
b_i''
 &= -\theta_i \,\,
    - \sum_{i+1\le j\le N} \frac{\delta_j L_{j-}}{1+\delta_j L_{j-}}\eta_{ij}
 \notag \\
 &\quad -\sum_{i+1\le k<l\le N}
         \frac{\delta_k L_{k-}}{1+\delta_k L_{k-}}
         \frac{\delta_l L_{l-}}{1+\delta_l L_{l-}}\zeta_{ikl},
\end{align}
where
\begin{align}
\theta_i = \kappa(\lambda_i),
\quad\text{}\quad
\eta_{ij} = \kappa(\lambda_i+\lambda_j)-\kappa(\lambda_i)-\kappa(\lambda_j)
\end{align}
and
\begin{align}
\zeta_{ikl}
 &= \widehat\kappa(\lambda_i+\lambda_k+\lambda_l)
 - \widehat\kappa(\lambda_i+\lambda_k) - \widehat\kappa(\lambda_i+\lambda_l)
 \notag\\ &\quad - \widehat\kappa(\lambda_k+\lambda_l)
 + \widehat\kappa(\lambda_i) + \widehat\kappa(\lambda_k)
 + \widehat\kappa(\lambda_l).
\end{align}
The number of terms to be calculated is thus reduced from $O(2^N)$ to $O(N^2)$,
while the error induced is
\begin{align}
b_i = b_i'' + O(N^2\delta^3\|L\|^3).
\end{align}
Therefore, the gain in computational time is significant, while the loss in
accuracy is usually relatively small. The numerical examples verify this, see
section \ref{drift-numerics} for more details.

\textbf{2.}
The second step is to approximate the random terms
\begin{align}
Z_j(t)
 := \frac{\delta_j L_j(t)}{1+\delta_j L_j(t)}
\quad\text{ and }\quad
\ykl(t)
 := \frac{\delta_k L_k(t)}{1+\delta_k L_k(t)}
    \frac{\delta_l L_l(t)}{1+\delta_l L_l(t)}
\end{align}
in \eqref{drift-2} by a \tih \lev process. Define the functions
\begin{align*}
f(x) = \frac{\delta_j \e^x}{1+\delta_j \e^x}
\quad\text{ and }\quad
g(x_k,x_l) = \frac{\delta_k \e^{x_k}}{1+\delta_k \e^{x_k}}
         \frac{\delta_l \e^{x_l}}{1+\delta_l \e^{x_l}},
\end{align*}
where
\begin{align*}
f'(x) = \frac{\delta_j \e^x}{(1+\delta_j \e^x)^2}
\quad\text{and}\quad
f''(x) = \frac{\delta_j \e^x(1-\delta_j \e^x)}{(1+\delta_j \e^x)^3}.
\end{align*}
The partial derivatives of $g$ can be computed equally easily, and are denoted
\begin{align}
g_k = \frac{\partial}{\partial x_k}g, \quad
g_l = \frac{\partial}{\partial x_l}g, \quad
g_\kl = \frac{\partial^2}{\partial x_k\partial x_l}g,
\end{align}
and so forth. We obviously have that
\begin{align}
Z_j(t) = f\big(G_j(t)\big)
\quad\text{and}\quad
Y_\kl(t) = g\big(G_k(t),G_l(t)\big).
\end{align}

The functions $f$ and $g$ are $C^2$-differentiable, hence we can apply It\^o's
formula for semimartingales (cf. e.g. \cite[Theorem I.4.57]{JacodShiryaev03})
to $Z_j$ and $\ykl$. Using \eqref{intL} we may derive (with time variable $s$
suppressed or denoted by $\cdot$ in the integrands)
\begin{align}\label{dZ}
\ud Z_{j}
 & = \bigg( \int_{\mathbb{R}^{m}}\left( f(G_{j}+\lambda_{j}^{\mathsf{T}}x)
     - f(G_{j})-f^{\prime}\left(G_{j}\right)
      \lambda_{j}^{\mathsf{T}}x\right)  F(\cdot,\dx)\\
 &\qquad + f^{\prime}\left( G_{j}\right)  b_{j}''
   + \frac{1}{2}f^{\prime\prime}\left(G_{j}\right)
     \left\vert \lambda_{j}\right\vert^{2}\alpha \bigg)\ds
   + f^{\prime}\left(G_{j}\right)
     \sqrt{\alpha}\lambda_{j}^{\mathsf{T}}\ud W\notag\\
 &\quad + \int_{\mathbb{R}^{m}}
   \left(f(G_{j-}+\lambda_{j}^{\mathsf{T}}x)-f(G_{j-})\right)
     \left(\mu\dsdx-F(\cdot,\dx)\ds\right).
\notag
\end{align}
The derivation is given in Appendix~\ref{DerZ}. Hence, we have that
\begin{eqnarray}\label{SDEZ}
\ud Z_{j}(s)
 &=& A_{j}(s,L(s))\ds
  + B_{j}^{\mathsf{T}}(s,L_{j}(s))\ud W(s) \notag\\
 &&+ \int_{\mathbb{R}^{m}}C_{j}(s,L_{j}(s),x)
       \left(\mu\dsdx-F(\cdot,\dx)\ds\right),
\end{eqnarray}
with obvious definitions of the {\em deterministic} functions $A_j,$ $B_j,$ and
$C_j$. Due to the drift term $b_{j}''$, the function $A_{j}$ depends on the
whole \lib vector $L$ rather than $L_j$ only.

Similarly, we have for \ykl that
\begin{eqnarray}\label{SDEY}
\ud \ykl(s)
 &=& A_\kl(s,L(s))\ds
  + B_\kl^{\mathsf{T}}(s,L_\kl(s))\ud W(s) \notag\\
 &&+ \int_{\mathbb{R}^{m}}C_\kl(s,L_\kl(s),x)
       \left(\mu\dsdx-F(\cdot,\dx)\ds\right),
\end{eqnarray}
where $A_\kl,$ $B_\kl,$ and $C_\kl$ are deterministic functions; see Appendix
\ref{DerY} for all the details. Analogously to \eqref{SDEZ}, $A_\kl$ depends on
the whole \lib vector $L$, while $B_\kl$ and $C_\kl$ depend on $L_k$ and $L_l$
only; this is denoted by $L_\kl$.

\textbf{3.}
The next step is to approximate $Z_j$ and \ykl by suitable \lev processes. This
approximation is based on a Picard iteration for the SDEs in \eqref{SDEZ} and
\eqref{SDEY}. Regarding $Z$, the initial value of the Picard iteration is
\begin{align}\label{Pic-1}
Z_j^{(0)}
 = Z_j(0)
 = \frac{\delta_j L_j(0)}{1+\delta_j L_j(0)},
\end{align}
while the first order Picard iteration is provided by
\begin{align}\label{Pic-2}
Z_{j}^{(1)}(t)
 &= Z_{j}(0) + \int_0^t A_{j}(s,L(0))\ds
  + \int_0^t B_{j}^{\mathsf{T}}(s,L_{j}(0))\ud W(s)\notag\\
 &\quad+ \int_0^t\int_{\mathbb{R}^{m}}C_{j}(s,L_{j}(0),x)
     \left(\mu\dsdx-F(\cdot,\dx)\ds\right).
\end{align}
We can easily deduce that $Z^{(1)}$ is a \textit{\tih L\'evy process}, since the
coefficients $A_j(\cdot,L(0)),$ $B_j(\cdot,L_j(0)),$ and
$C_j(\cdot,L_j(0),\cdot)$ in \eqref{Pic-2} are \emph{deterministic}. Indeed, we
have that
\begin{multline}\label{Pic-A}
A_{j}(s,L(0))
 = f^{\prime}\left(G_{j}(0)\right) b_{j}^{(0)}(s)
 + \frac{1}{2}f^{\prime\prime}\left(G_{j}(0)\right)
    \left\vert \lambda_{j}\right\vert ^{2}(s)\alpha(s) \\
 + \int_{\mathbb{R}^{m}}\left(
      f(G_{j}(0)+\lambda_{j}^{\mathsf{T}}(s)x)-f(G_{j}(0))
       - f^{\prime}\left(G_{j}(0)\right) \lambda_{j}^{\mathsf{T}}(s)x\right)
        F(\cdot,\dx),
\end{multline}
where
\begin{align*}
b_{j}^{(0)}(s)
 &:= -\theta_i(s) \,\,
    - \sum_{i+1\le j\le N}
       \frac{\delta_j L_{j-}(0)}{1+\delta_j L_{j-}(0)}\eta_{ij}(s)  \notag \\
 &\quad -\sum_{i+1\le k<l\le N}
         \frac{\delta_k L_{k-}(0)}{1+\delta_k L_{k-}(0)}
         \frac{\delta_l L_{l-}(0)}{1+\delta_l L_{l-}(0)}\zeta_{ikl}(s),
\end{align*}
and
\begin{align}\label{Pic-B} 
B_{j}(s,L_{j}(0))
 &= f^{\prime}\left(G_{j}(0)\right)
    \sqrt{\alpha(s)}\lambda_{j}(s),\\ \label{Pic-C}
C_{j}(s,L_{j}(0),x)
 &= f\big(G_{j}(0)+\lambda_{j}^{\mathsf{T}}(s)x\big)-f(G_{j}(0)).
\end{align}

Analogously, the initial value of the Picard iteration for \eqref{SDEY} is
\begin{align}
\ykl^{(0)}
 =\ykl(0)
 = \frac{\delta_k L_k(0)}{1+\delta_k L_k(0)}
    \frac{\delta_l L_l(0)}{1+\delta_l L_l(0)},
\end{align}
and the first order iteration is
\begin{align}\label{Pic-2-Y}
\ykl^{(1)}(t)
 &= \ykl(0) + \int_0^t A_\kl(s,L(0))\ds
  + \int_0^t B_\kl^{\mathsf{T}}(s,L_\kl(0))\ud W(s) \notag\\
 &\quad+ \int_0^t\int_{\mathbb{R}^{m}}C_\kl(s,L_\kl(0),x)
       \left(\mu\dsdx-F(\cdot,\dx)\ds\right),
\end{align}
and we can again deduce that $\ykl^{(1)}$ is an additive \lev process.

\textbf{4.}
The fourth step is to apply the \lev approximations of the random terms to
\eqref{drift-2}. Let us denote by $\widehat{b}_{i}$ the resulting approximate
drift term; we have that
\begin{align}\label{st}
b_{i}''
 \approx \widehat{b}_{i}
 &:= -\theta_{i} \,\, - \sum_{i+1\le j\le N}\eta_{ij}Z_{j}^{(1)}
   - \sum_{i+1\le k<l\le N}\zeta_{ikl} \ykl^{(1)}.
\end{align}
Keeping in mind that $\widehat{b}_{i}$ will be integrated over time, we define
\begin{align*}
V_{ij}(s,t)=\int_{s}^{t}\eta_{ij}(r)\ud r,
 \quad\text{and}\quad
\overline{V}_{ikl}(s,t)=\int_{s}^{t}\zeta_{ikl}(r)\ud r,
\end{align*}
which are obviously deterministic processes of finite variation. Now, for fixed
$t>0$, we can apply integration by parts, which yields
\begin{align}\label{IBP-1}
\int_{0}^{t}\eta_{ij}(s)Z_{j}^{(1)}(s)\ds
 &\stackrel{\eqref{Pic-2}}{=}
   V_{ij}(0,t)Z_j(0) + \int_{0}^{t}V_{ij}(s,t) A_{j}(s,L(0))\ds \nonumber\\
 &\qquad + \int_{0}^{t}V_{ij}(s,t) B_{j}^{\mathsf{T}}(s,L_{j}(0))\ud W(s) \\
 &\qquad + \int_{0}^{t}V_{ij}(s,t)  \int_{\mathbb{R}^{m}} 
             C_{j}(s,L_{j}(0),x) \mt\dsdx.\nonumber
\end{align}
Similarly for the other term we get
\begin{align}\label{IBP-2}
\int_{0}^{t}\zeta_{ikl}(s)Y_\kl^{(1)}(s)\ds
 &\stackrel{\eqref{Pic-2-Y}}{=} 
  \oV_{ikl}(0,t)Y_\kl(0) + \int_{0}^{t} \oV_{ikl}(s,t) A_\kl(s,L(0))\ds 
  \nonumber\\
 &\qquad  + \int_{0}^{t} \oV_{ikl}(s,t) B_\kl^{\mathsf{T}}(s,L_\kl(0))\ud W(s)\\
 &\qquad + \int_{0}^{t} \oV_{ikl}(s,t)  \int_{\mathbb{R}^{m}}
     C_\kl(s,L_\kl(0),x) \mt\dsdx.\nonumber
\end{align}

\textbf{5.}
Finally, collecting all the pieces together we can derive a \lev approximation
for the log-\lib rates. The \textit{approximate} log-\lib is denoted by
$\widehat{G}_i$ and has the following dynamics
\begin{align}\label{approx-log-LIBOR}
\widehat{G}_i(t)
 &= G_i(0) + \int_{0}^{t}\widehat{b}_{i}(s)\ds
  + \int_{0}^{t}\lambda_{i}^{\mathsf{T}}(s)\ud H(s),
\end{align}
which using \eqref{st}, \eqref{IBP-1} and \eqref{IBP-2} leads to
\begin{align}\label{col}
\widehat{G}_i(t)
 &= \widehat{G}_i(0,t) + \int_{0}^{t} \mathrm{H}_i(t,s)\ds
  + \int_{0}^{t}\Theta_i^{\mathsf{T}}(t,s)\ud W(s)
  + \int_{0}^{t} \mathrm{I}_i(t,s,x)\mt\dsdx,
\end{align}
where
\begin{align*}
\widehat{G}_i(0,t)
 &:= {G}_i(0) - \sum_{i+1\le j\le N} V_{ij}(0,t)Z_j(0)\\
 &\quad - \sum_{i+1\le k<l\le N}\oV_{ikl}(0,t)Y_\kl(0), \\
\mathrm{H}_i(t,s)
 &:= - \theta_{i}(s) 
  - \sum_{i+1\le j\le N}V_{ij}(s,t) A_{j}(s,L(0))\\
 &\quad - \sum_{i+1\le k<l\le N} \oV_{ikl}(s,t) A_\kl(s,L(0)),
\end{align*}
\begin{align*}
\Theta_i^{\mathsf{T}}(t,s) 
 &:= \sqrt{\alpha(s)}\lambda_{i}^{\mathsf{T}}(s)
   -\sum_{i+1\le j\le N}V_{ij}(s,t) B_{j}^{\mathsf{T}}(s,L_{j}(0))\\
 &\quad - \sum_{i+1\le k<l\le N} \oV_{ikl}(s,t) B_\kl^{\mathsf{T}}(s,L_\kl(0))
\end{align*}
and
\begin{align*}
\mathrm{I}_i(t,s,x)
 &:= \lambda_{i}^{\mathsf{T}}(s)x
   -\sum_{i+1\le j\le N}V_{ij}(s,t) C_{j}(s,L_{j}(0),x)\\
 &\quad-\sum_{i+1\le k<l\le N} \oV_{ikl}(s,t) C_\kl(s,L_\kl(0),x).
\end{align*}
Let us introduce the process $X_i^{(t)}(r),$ $0\leq r\leq t$, defined by
\begin{align*}
X_i^{(t)}(r) \!
 := \widehat{G}_i(0,r) +\! \int_{0}^{r} \mathrm{H}_i(t,s)\ds
  +\! \int_{0}^{r}\Theta_i^{\mathsf{T}}(t,s)\ud W(s)
  +\! \int_{0}^{r}\mathrm{I}_i(t,s,x)\mt\dsdx.
\end{align*}
Obviously, $X_i^{(t)}(r)$, $0\le r\le t$ is a \tih \lev process whose
characteristic function may be expressed by the \lk formula in terms of
$\mathrm{H}_i$, $\Theta_i$ and $\mathrm{I}_i$ in a straightforward manner.

\begin{remark}
We will call the approximation in \eqref{col} the \textit{second order log-\lev
approximation} of the \lib rate. If we ignore the second order terms (i.e. those
depending on $L_k$ and $L_l$), we immediately arrive at the \textit{first order
approximation}. The numerical results in section \ref{numerics} document the
improvement from the first to the second order approximation.
\end{remark}

\begin{remark}\label{DL-KSS}
If we restrict our model to the Brownian motion case, the approximation in
\eqref{col} coincides with the ``fully lognormal model'' of
\cite{DanilukGatarek05}; see also
\cite{KurbanmuradovSabelfeldSchoenmakers}.
\end{remark}

\begin{remark}
Note that the approximation methods developed in the previous sections do not
depend crucially on the choice of the measure. If we work under the spot
measure, cf. Remark \ref{remP}, then the Picard approximations can be carried
out similarly. However, an additional approximation is required to represent the
drift in terms of cumulants as in eq. \eqref{drift-effi} (because of the
$1/\prod_j$ terms).
\end{remark}

\subsection{Efficient simulation of the log-\lev approximation}
\label{eff-sim}

In this section, we outline how simulation of the L\'evy approximation
\begin{equation}\label{addp}
\widehat{G}_{i}(t)
 = \widehat{G}_{i}(0,t) + \int_{0}^{t}\mathrm{H}_{i}(t,s)\ds
 + \int_{0}^{t}\Theta_{i}^{\mathsf{T}}(t,s)\ud W(s)
 + \int_{0}^{t}\mathrm{I}_{i}(t,s,x)\mt\dsdx
\end{equation}
can be carried out in an effective way due to the fact that 
$\widehat{G}_{i}(0,t)$ and the integrands in \eqref{addp} are explicitly known
deterministic functions.\\

\noindent \textbf{(I)}
The terms $\widehat{G}_{i}(0,t)$ and $\int_{0}^{t}\mathrm{H}_{i}(t,s)\ds$ are
deterministic integrals which may be computed outside any Monte Carlo
loop using some quadrature formula.\\

\noindent \textbf{(II)}
The Gaussian part
\begin{equation}\label{G}
\varsigma_{i}(t):=\int_{0}^{t}\Theta_{i}^{\mathsf{T}}(t,s)\ud W(s)
\end{equation}
may be computed either by usual Euler stepping, or even directly at some fixed
time $t$ if only the distribution of $\widehat{G}(t)$ matters. In this respect,
the distribution of any vector $(\varsigma_{i_{1}}(t),...,\varsigma_{i_{k}}(t))$
--- for simulating a set of log-LIBORs
$(\widehat{G}_{i_{1}}(t),...,\widehat{G}_{i_{k}}(t))$) --- is Gaussian with
explicitly known covariance structure, and thus can be simulated
straightforwardly.\\

\noindent \textbf{(III)}
Finally, consider the practically important case where the L\'evy measure itself
is time homogeneous, i.e. $F(\dx)\equiv F(\cdot,\dx)$. After truncating this
measure with respect to jumps with size smaller than some $\epsilon>0$ (if
needed), simulation of a realization of the jump term in \eqref{addp} may
effectively be carried out as follows. First sample on the interval $(0,t)$ the
number $N_{t}$  (of jump times) according to a Poisson distribution with
intensity $tF(\left\{  ||x||>\epsilon\right\} ).$ Next distribute $N_{t}$
jump points $\left\{s_{1},...,s_{N_{t}}\right\}$ uniformly over the interval
$(0,t),$ and sample independently for each jump point $s_{l}$ a jump $x_{l},$
$1\leq l\leq N_{t}$ from the probability measure 
\[
\frac{F(\dx\cap\left\{||x||>\epsilon\right\})}
     {F(\left\{||x||>\epsilon\right\})}.
\]
Then a realization of the (compensated) jump term is obtained as
\begin{equation}\label{J}
\varsigma_{i}^{J}(t)
 := \sum_{l=1}^{N_{t}}\mathrm{I}_{i}(t,s_{l-},x_{l})
  - \int_{0}^{t}\int_{||x||>\epsilon}\mathrm{I}_{i}(t,s,x)F(\dx)\ds,
\end{equation}
where the deterministic integral term can be computed outside any Monte Carlo
loop by standard methods. Note that a realization of the whole log-\lib vector
$(\varsigma_{1}^{J}(t),\dots,\varsigma_{N}^{J}(t))$ will be computed using the
same set of jumps $(s_{l},x_{l}),$ $l=1,...,N_{t}$.\\

The main benefit from the log-L\'evy approximation as outlined above, is the
fact that for the simulation of a log-LIBOR vector
$(\widehat{G}_{i}(t),...,\widehat{G}_{N}(t))$, the computation of the terms in
\eqref{logl} via \eqref{drift-effi} or \eqref{drift-2} based on each realization
of the Brownian motion and the jump process on a fine enough time grid is
\textit{not} required. This is in clear contrast to the Euler (or
predictor-corrector) discretization of \eqref{logl} and \eqref{drift-effi}. It
is obvious that in view of the complex structure of \eqref{drift-2} only, such a
simulation would require the (accurate enough) construction of a whole log-LIBOR
system $(\widehat{G}_{i}(t_{j}),...,\widehat{G}_{N}(t_{j}))$ for $0<t_{1}<\cdot
\cdot\cdot<t_{n}:=t$  involving the evaluation of the function
$b^{\prime\prime}$ at each grid point $t_{j}.$ In contrast, simulation of the
log-L\'evy LIBOR approximation only involves the evaluation of \eqref{J} at the
jump times and the relatively efficient simulation of the Wiener integral
\eqref{G} inside a Monte Carlo loop.

\section{Error estimates}
\label{error}

In this section, we will provide some error estimates for the log-\lev
approximations in order to offer a theoretical justification for the proposed
approximations. The error estimates are rather qualitative in nature, however
they allow for useful conclusions.

In view of \eqref{approx-log-LIBOR} we have for the pathwise error of the
(log-)LIBOR approximation,
\begin{align*}
\left\vert \frac{\widehat{L}_{i}(t)}{L_{i}(t)} \right\vert
 \leq \exp\left\vert\widehat{G}_{j}(t)-G_{j}(t)\right\vert
 \leq \exp\left(\int_{0}^{t}
       \left\vert\widehat{b}_{i}(s)-b_{i}(s)\right\vert \ds\right),
\end{align*}
thus we need to study the difference $|\widehat{b}_{i}-b_{i}|$. Since the main
contribution of this error is due to the first and second order term in (2.7),
we consider instead (see \eqref{drift-2})
\[
\left\vert \widehat{b}_{i} - b_{i}^{\prime\prime}\right\vert
\leq\sum_{i+1\leq j\leq N}\left\vert Z_{j}-Z_{j}^{(1)}\right\vert \left\vert
\eta_{ij}\right\vert +\sum_{i+1\leq k<l\leq N}\left\vert Y_{kl}-Y_{kl}%
^{(1)}\right\vert \left\vert \zeta_{ikl}\right\vert.
\]
Let us assume for simplicity that $\alpha(s)\equiv1,$ and that $K_{\eta}$ and
$K_{\zeta}$ are (dimensionless) constants such that
\begin{align*}
\max_{1\leq i<j\leq N}\left\vert \eta_{ij}\right\vert  &  \leq K_{\eta}%
\max_{1\leq i\leq N}\sup_{0\leq t\leq T}\left\Vert \lambda_{i}(t)\right\Vert
_{2}^{2}=:K_{\eta}\lambda_{\max}^{2},\\
\max_{1\leq i<k<l\leq N}\left\vert \zeta_{ikl}\right\vert  &  \leq K_{\zeta
}\max_{1\leq i\leq N}\sup_{0\leq t\leq T}\left\Vert \lambda_{i}(t)\right\Vert
_{2}^{2}=:K_{\zeta}\lambda_{\max}^{2}.
\end{align*}
We then have
\begin{gather*}
\left\Vert \log\left\vert \frac{\widehat{L}_{i}(t)}{L_{i}(t)}\right\vert
\right\Vert _{L_{2}(\mathbb{P}_{\ast})}\leq K_{\eta}\lambda_{\max}^{2}%
\max_{i+1\leq j\leq N}\int_{0}^{t}\left\Vert Z_{i}^{(1)}(s)-Z_{i}%
(s)\right\Vert _{L_{2}(\mathbb{P}_{\ast})}\mathrm{d}s\\
+K_{\zeta}\lambda_{\max}^{2}\max_{i+1\leq k<l\leq N}\int_{0}^{t}\left\Vert
Y_{kl}^{(1)}(s)-Y_{kl}(s)\right\Vert _{L_{2}(\mathbb{P}_{\ast})}%
\mathrm{d}s=:(I)+(II).
\end{gather*}
For the term $(I)$ we get from \eqref{SDEZ} and \eqref{Pic-2}
\begin{multline*}
\left\Vert Z_{j}^{(1)}(s)-Z_{j}(s)\right\Vert _{L_{2}(\mathbb{P}_{\ast})} 
\leq\int_{0}^{s}\left\vert A_{j}(u,L(0))-A_{j}(u,L(u))\right\vert
_{L_{2}(\mathbb{P}_{\ast})}\mathrm{d}u
\\  +\left(  \int_{0}^{s}E\left\Vert B_{j}(u,L_{j}(0))-B_{j}(u,L_{j}%
(u))\right\Vert _{2}^{2}\mathrm{d}u\right)  ^{1/2}
\\ +\left(  \int_{0}^{s}\int_{\mathbb{R}^{m}}E\left(  C_{j}(u,L_{j}%
(0),x)-C_{j}(u,L_{j}(u),x)\right)  ^{2}F(u,\mathrm{d}x)du\right)  ^{1/2}.
\end{multline*}
In view of \eqref{Pic-A}, \eqref{Pic-B} and \eqref{Pic-C}, let $K_{A},$ $K_{B},$
$K_{C}$ be dimensionless Lipschitz constants such that for all $1\leq j\leq N$
and $0\leq u\leq T_{\ast},$%
\begin{align*}
\left\vert A_{j}(u,y)-A_{j}(u,y^{\prime})\right\vert  &  \leq K_{A}%
\lambda_{\max}^{2}\left\Vert y-y^{\prime}\right\Vert _{2},\\
\left\Vert B_{j}(u,y_{j})-B_{j}(u,y_{j}^{\prime})\right\Vert _{2} &  \leq
K_{B}\lambda_{\max}\left\vert y_{j}-y_{j}^{\prime}\right\vert ,\\
\int_{\mathbb{R}^{m}}\left(  C_{j}(u,y_{j},x)-C_{j}(u,y_{j}^{\prime
},x)\right)  ^{2}F(u,\mathrm{d}x) &  \leq K_{C}^{2}\lambda_{\max}%
^{2}\left\vert y_{j}-y_{j}^{\prime}\right\vert ^{2}.
\end{align*}
Then, using
\begin{multline*}
\left\Vert Z_{j}^{(1)}(s)-Z_{j}(s)\right\Vert _{L_{2}(\mathbb{P}_{\ast})} 
\leq K_{A}\lambda_{\max}^{2}\int_{0}^{s}\left\Vert L(0)-L(u)\right\Vert
_{2,L_{2}(\mathbb{P}_{\ast})}\mathrm{d}u\\
+\left(  K_{B}+K_{C}\right)  \lambda_{\max}\left( \int_{0}^{s}
 E\left\vert L_{j}(0))-L_{j}(u)\right\vert ^{2}\mathrm{d}u\right)^{1/2},
\end{multline*}
we obtain the estimate%
\begin{multline*}
(I)\leq
 \lambda_{\max}^{4}K_{\eta}K_{A}\int_{0}^{t} \left( \int_{0}^{s}
  \left\Vert L(0)-L(u) \right\Vert_{2,L_{2}(\mathbb{P}_{\ast})}\mathrm{d}u
  \right) \ds\\
 + \lambda_{\max}^{3} K_{\eta} \left(K_{B}+K_{C}\right) 
    \int_{0}^{t} \max_{i+1\leq j\leq N} \left( \int_{0}^{s}
   E\left\vert L_{j}(0))-L_{j}(u) \right\vert ^{2}\mathrm{d}u \right)^{1/2} \ds,
\end{multline*}
and a similar expression may be obtained for the second term $(II).$

On an intuitive level we may interpret the estimates $(I)$ and $(II)$ in the
following way: if we roughly consider that (the approximate squared variance)
$E\left\vert L_{j}(0))-L_{j}(u)\right\vert ^{2}\lessapprox \lambda_{\max}^{2}u,$
then for $(I)$ we obtain
\begin{align*}
(I) 
 &\lessapprox \lambda_{\max}^{5}K_{\eta}K_{A} \int_{0}^{t} \int_{0}^{s}
               \sqrt{u}\,\ud u\, \ds\\
 &\quad+ \lambda_{\max}^{4}K_{\eta} \left(K_{B}+K_{C}\right)  \int_{0}^{t}
     \left(\int_{0}^{s}u\mathrm{d}u\right)^{1/2}\ds \\
 &= \frac{4}{15}\lambda_{\max}^{5}K_{\eta}K_{A}t^{5/2}
  + \frac{\sqrt{2}}{4}\lambda_{\max}^{4}K_{\eta}\left(K_{B}+K_{C}\right) t^{2},
\end{align*}
and a similar result for $(II).$ Hence, for some dimensionless constants $K_{1}$
and $K_{2},$
\[
\left\Vert \log\left\vert \frac{\widehat{L}_{i}(t)}{L_{i}(t)}\right\vert
\right\Vert _{L_{2}(\mathbb{P}_{\ast})}\lessapprox K_{1}\left(  \lambda_{\max
}^{2}t\right)  ^{5/2}+K_{2}\left(  \lambda_{\max}^{2}t\right)  ^{2}.
\]
Concluding, the log-L\'evy LIBOR approximations are extremely good as long as
$\lambda_{\max}^{2}t$ is small enough but, may become poor as soon as this
product grows very large. This issue is confirmed in our numerical experiments.

\section{Numerical illustrations}
\label{numerics}

Throughout this section, we will consider a simple example with a flat and
constant volatility structure. Similarly zero coupon rates are generated from a
flat term structure of interest rates: $B(0,T_i)=\exp(-0.04\cdot T_i)$. We
consider a tenor structure with 6 month increments (i.e. $\delta_i=\frac12$). As
stated in the introduction, the Brownian motion case is already well studied;
therefore we set $\alpha=0$, thus limiting ourselves to the case where $H$ is a
pure jump \lev process. We consider two univariate specifications, for
simplicity. The first is a tempered stable or CGMY process (cf.
\cite{Carretal02} and \cite{MadanYor08}) with parameters $M=G=13$, $Y=0.25$
and $C=48.4201$, resulting in a process with mean zero and variance 1 (at
$t=1$), infinite activity and finite variation. The CGMY process has cumulant
generating function defined for all $u\in\mathbb{C}$ with $|\Re u|\le\min(G,M)$,
\begin{align}
\label{kCGMY}
\kappa_{\text{CGMY}}(u)
 &= \Gamma(-Y)G^Y\left\{\left(1-\frac{u}{G}\right)^Y-1+\frac{uY}{G}\right\}
 \nonumber \\ &\quad
  + \Gamma(-Y)M^Y\left\{\left(1+\frac{u}{M}\right)^Y-1-\frac{uY}{M}\right\}.
\end{align}
The necessary conditions are then satisfied for term structures up to at least
10 years of length because $\overline{M}=\min(G,M)$, hence
$\sum_{i=1}^{20}|\lambda_i|\le12<\overline{M}$. Exact simulation of the
increments can be performed without approximation using the approach in
\cite{PoirotTankov06}. This approach can be used when simulating from
\eqref{intL} with or without drift expansions, but cannot be employed in the
case of the log-\lev approximation in \eqref{col} where jump sizes are
transformed in a non-linear fashion. Instead we employ an approximation where we
replace jumps smaller than $\epsilon$ with their expectation which is zero since
the jumps are compensated. This means that jumps bigger than $\epsilon$ follow a
compound Poisson process which can be easily simulated using the so-called
Rosinski rejection method (see \cite{Rosinski01} and
\cite[p.~338]{AsmussenGlynn07}). We set the truncation point sufficiently low,
at $\epsilon=10^{-3}$, thus making the variance of the truncated term
$\int_{-\epsilon}^{\epsilon}x^2\nu(dx)=3.11\times10^{-4}$, which can be
considered small enough to safely disregard. To be consistent, we employ
this procedure everywhere we simulate from the CGMY process.

The second specification is a compound Poisson process with normally distributed
jump sizes --- often referred to as the Merton model. The cumulant generating
function for $u\in\mathbb{C}$  is
\begin{align}
\kappa_{\text{Merton}}(u)= \bar\lambda\left(\exp (\bar\mu u +
\bar\sigma^2u^2)-1-\bar\mu u \right).
\end{align}
We set $\bar\lambda=5,\bar\mu=0$ and $ \bar\sigma=\sqrt{1/\bar\lambda}$ yielding
a process with mean zero and variance 1 (at $t=1$), as before.

In order to verify the validity of our approximations we consider linear,
nonlinear and path-dependent payoffs; in particular, forward rate agreements
(FRAs), caplets, swaptions and so-called sticky ratchet caplets. To price FRAs
and caplets  with strike $K$ maturing at time $T_i$, we compute the following
expectations: 
\begin{align}\label{cpl}
\mathbb{FRA}_0
 &= \delta_i B_{N+1}(0)\,\E_{\P_*}\Big[\prod_{l=i+1}^{N}
    \big(1+\delta_l L_l(T_{i+1})\big)(L_i(T_i)-K)\Big],\\
\mathbb{C}_0
 &= \delta_i B_{N+1}(0)\,\E_{\P_*}\Big[\prod_{l=i+1}^{N}
    \big(1+\delta_l L_l(T_{i+1})\big)(L_i(T_i)-K)^+\Big].
\end{align}
Following \cite[pp. 78]{Kluge05}, we have that the price of a payer swaption
with strike rate $K$, where the underlying swap starts at time $T_i$ and matures
at $T_m$ ($i<m\le N$) is given by
\begin{align}\label{swap-1}
\mathbb{S}_{0}
&= B_{N+1}(0)\, \E_{\P_*} \left[\left( - \sum^m_{k=i}
\bigg(c_k \prod_{l=k}^{N}\left(1+\delta_l L_l(T_i)\right)\bigg)\right)^+\right],
\end{align}
where
\begin{align}\label{ck}
c_k = \left\{
        \begin{array}{ll}
          -1,           & k=i,\\
          \delta_k K,   & \hbox{$i+1\le k\le m-1$,} \\
          1+\delta_k K, & \hbox{$k=m$.}
        \end{array}
      \right.
\end{align}
Similarly, a sticky ratchet caplet, which is a path-dependent derivative, can
priced by computing the following expectation:
\begin{align}
\mathbb{R}_0
 &= \delta_i B_{N+1}(0)\,\E_{\P_*}\Big[\prod_{l=i+1}^{N}
    \big(1+\delta_l L_l(T_{i+1})\big)(R_i(T_i))^+\Big],
\end{align}
where 
$$R_i(t)=L_i(t)-\min\{L_1(T_1),\dots,L_{i-1}(T_{i-1})\},\quad \forall t\in
[T_1,T_i].$$
Note that sticky ratchet caplets are often embedded in mortgages as a protection
against interest rates moving above a historical minimum value.

\subsection{Performance of the drift expansion}
\label{drift-numerics}

As we have argued in section \ref{effi}, the truncation of the drift term in
equation \eqref{dr} is necessary in order to build a model that is
computationally tractable. This section illustrates the effect of this
truncation using the standard Euler discretization of the actual dynamics, i.e.
equations \eqref{logl} and \eqref{drift-effi}. 

\begin{figure}[ht!]
 \centering
\includegraphics[width=6.25cm]{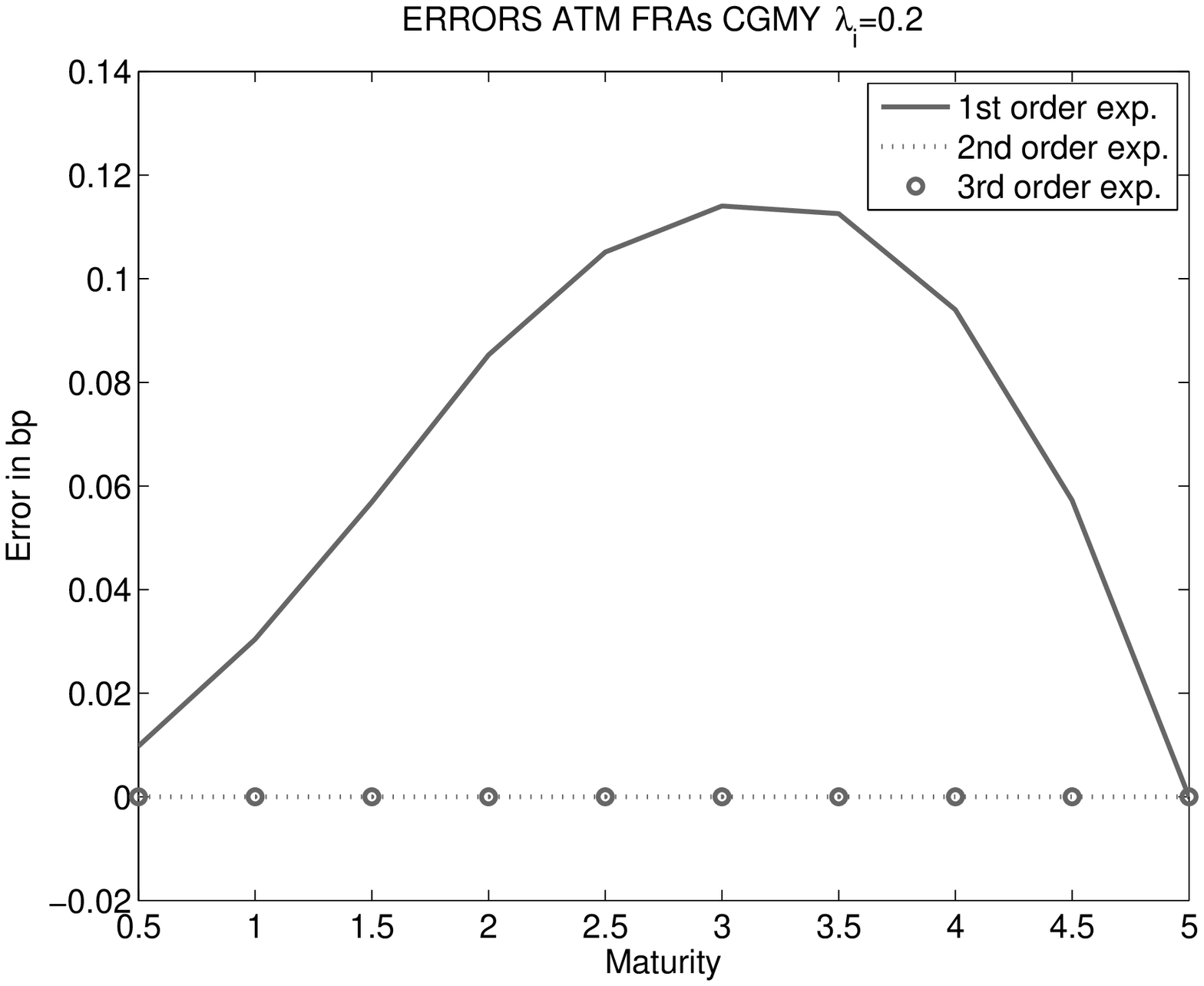}
\includegraphics[width=6.25cm]{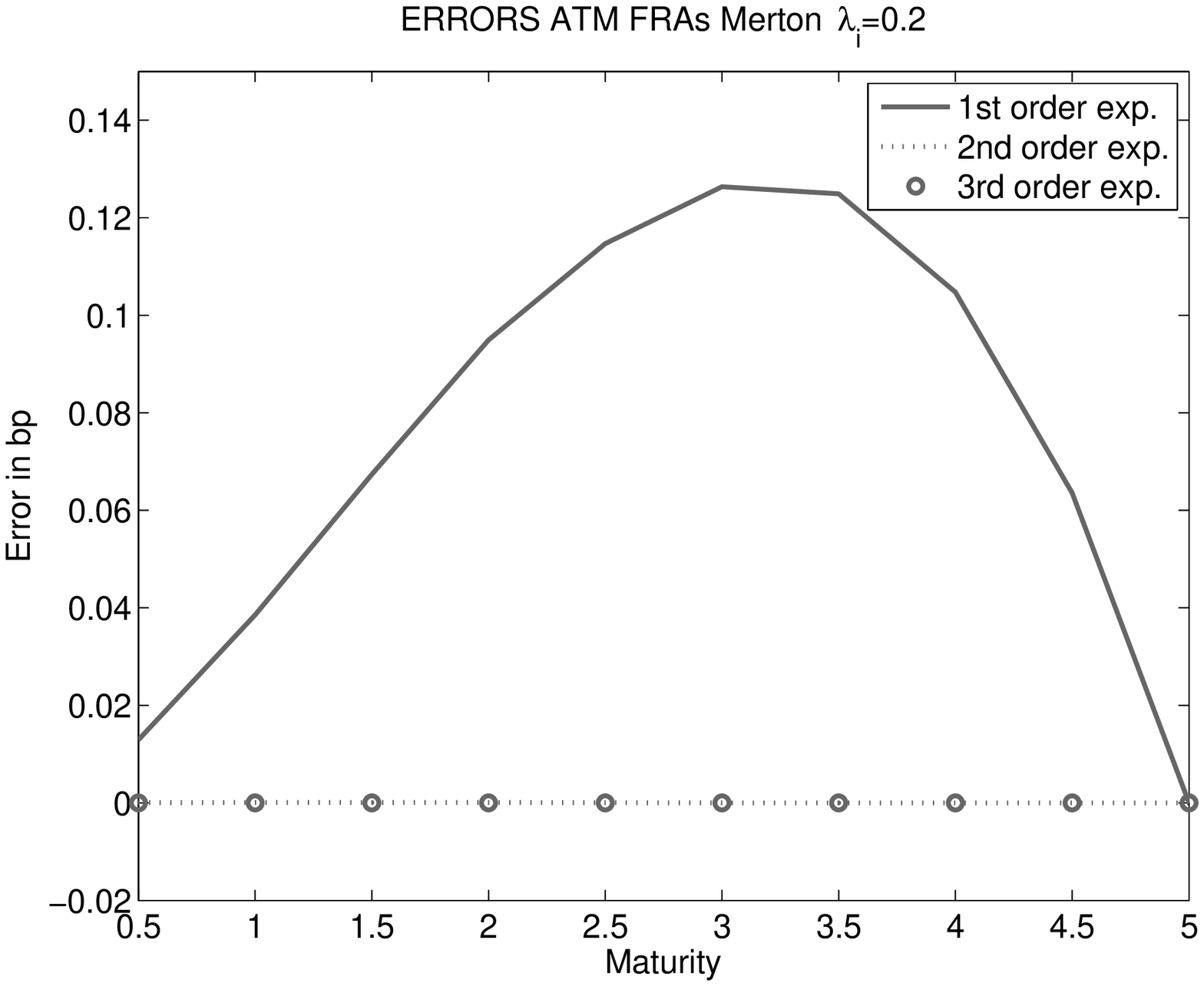}\\
\includegraphics[width=6.25cm]{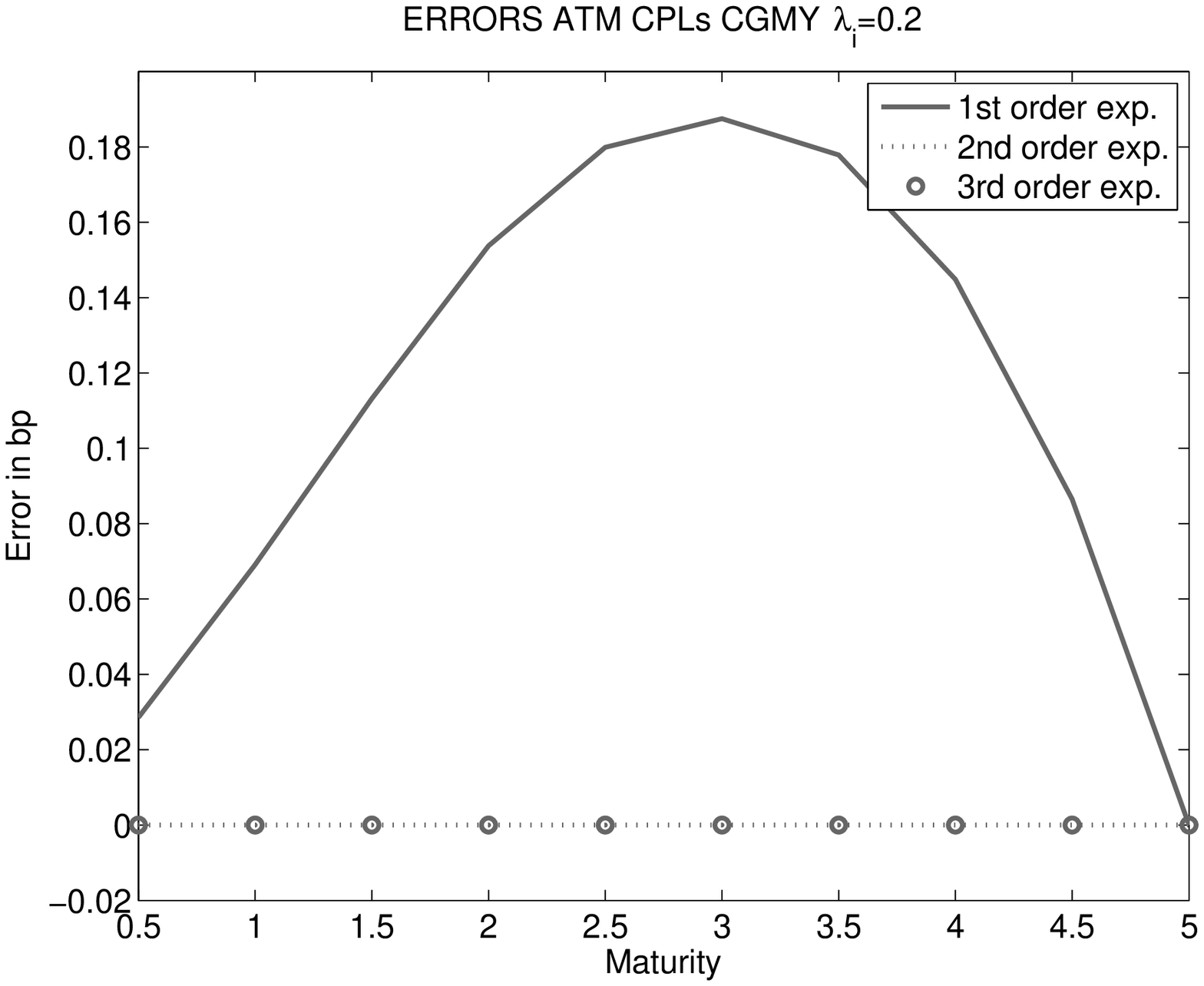}
\includegraphics[width=6.25cm]{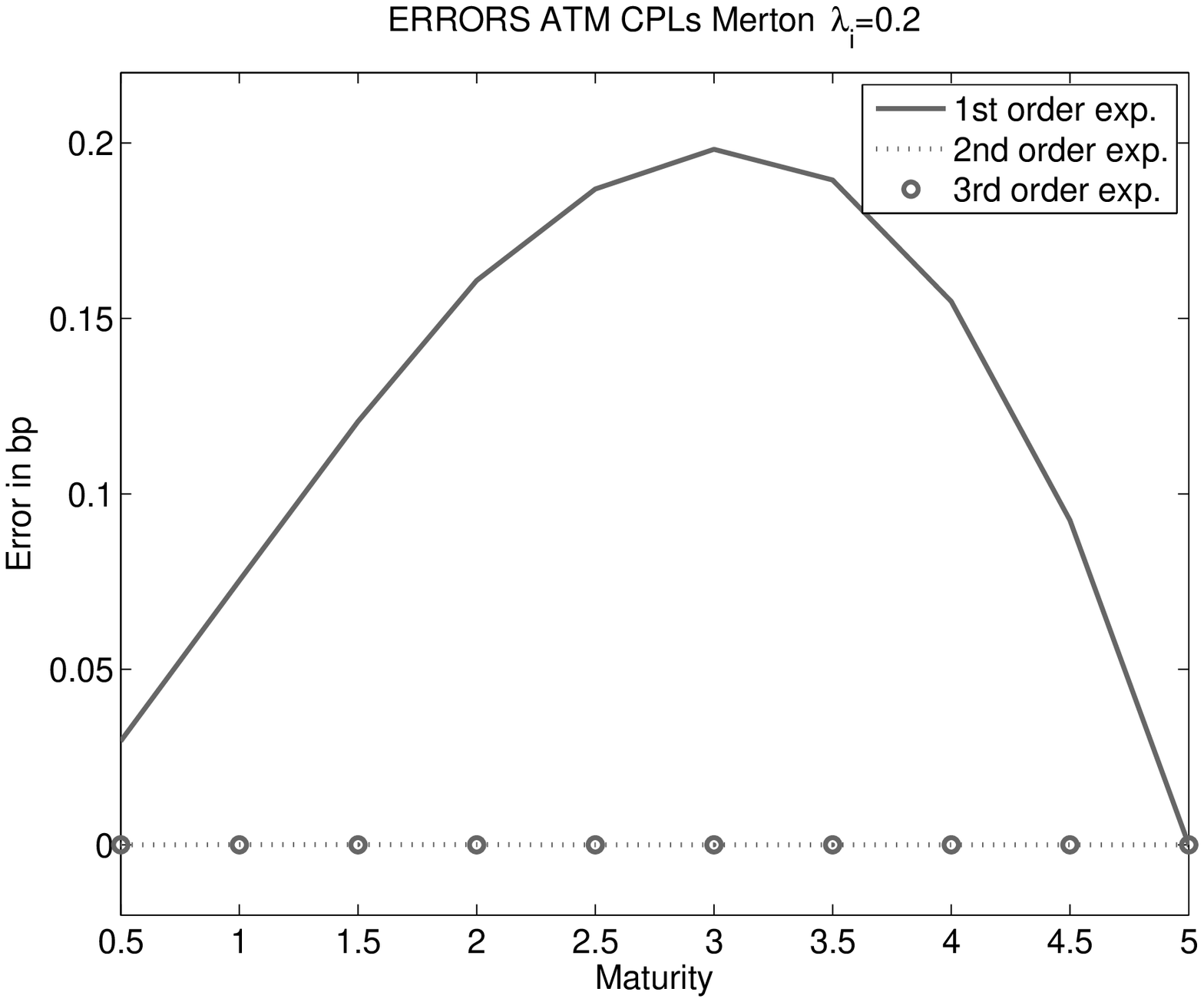}\\
 \caption{Drift expansion: low volatility scenario.}
 \label{fig:Drift-Low}
\end{figure}

\begin{figure}[ht!]
 \centering
\includegraphics[width=6.25cm]{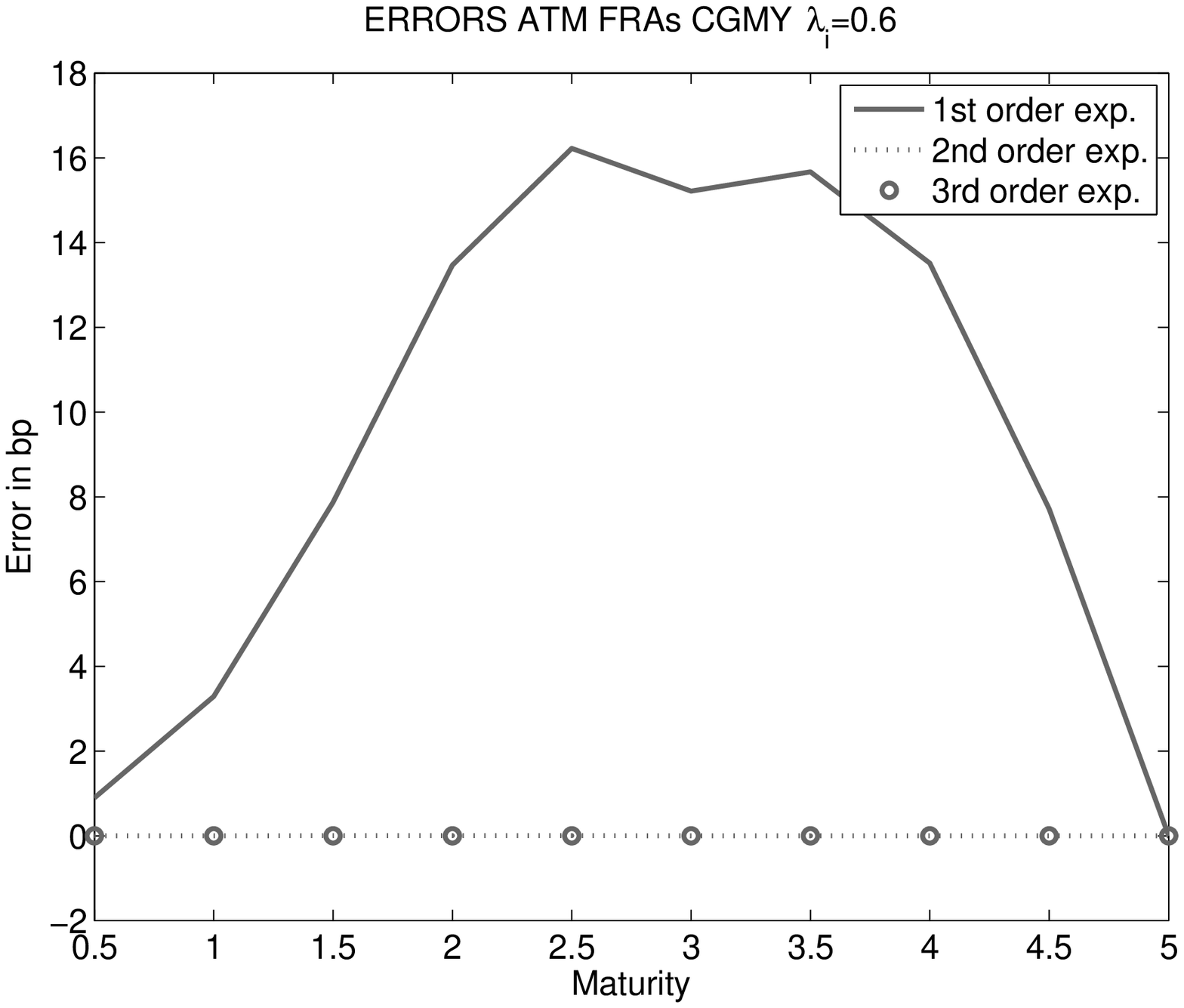}
\includegraphics[width=6.25cm]{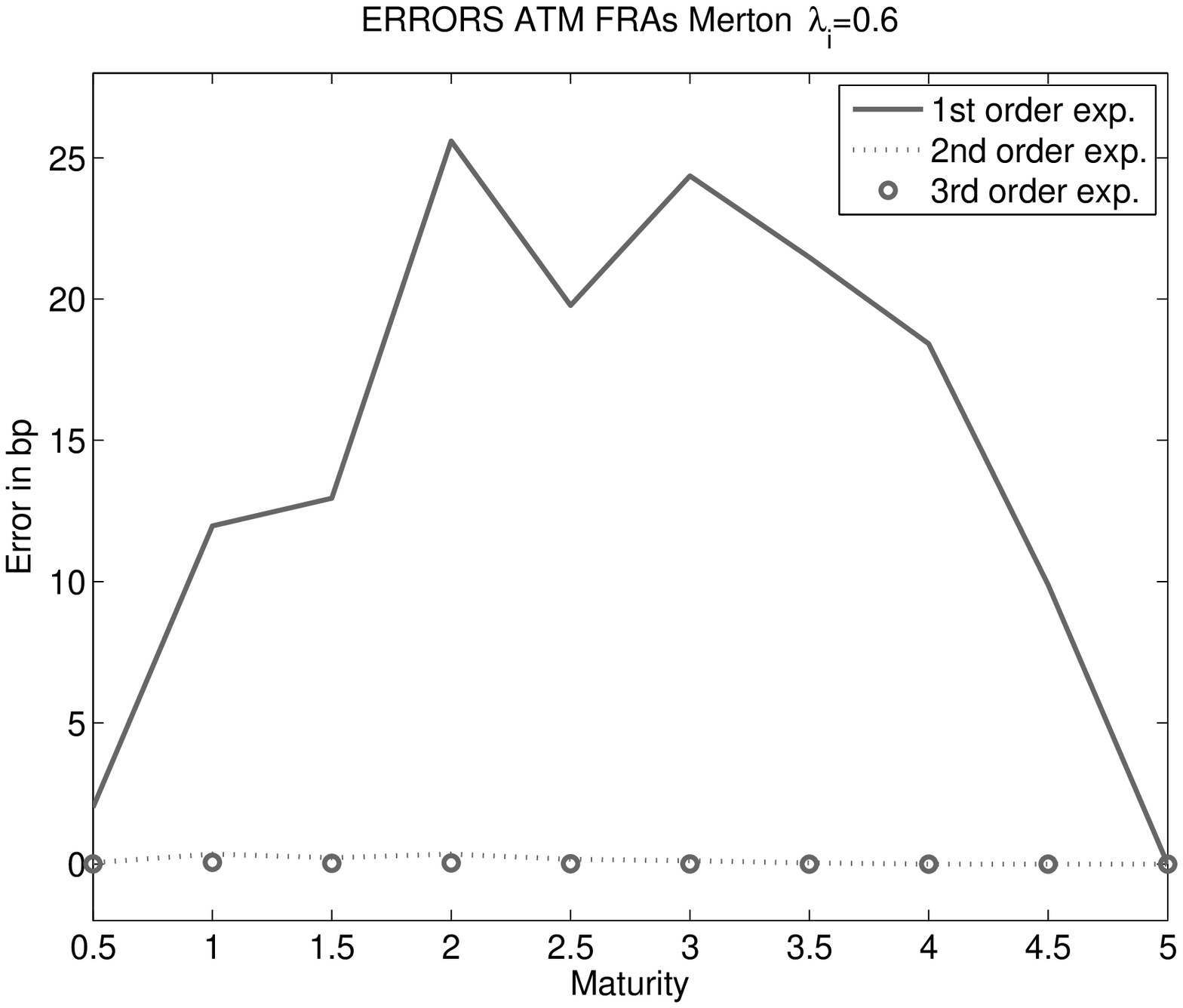}\\
\includegraphics[width=6.25cm]{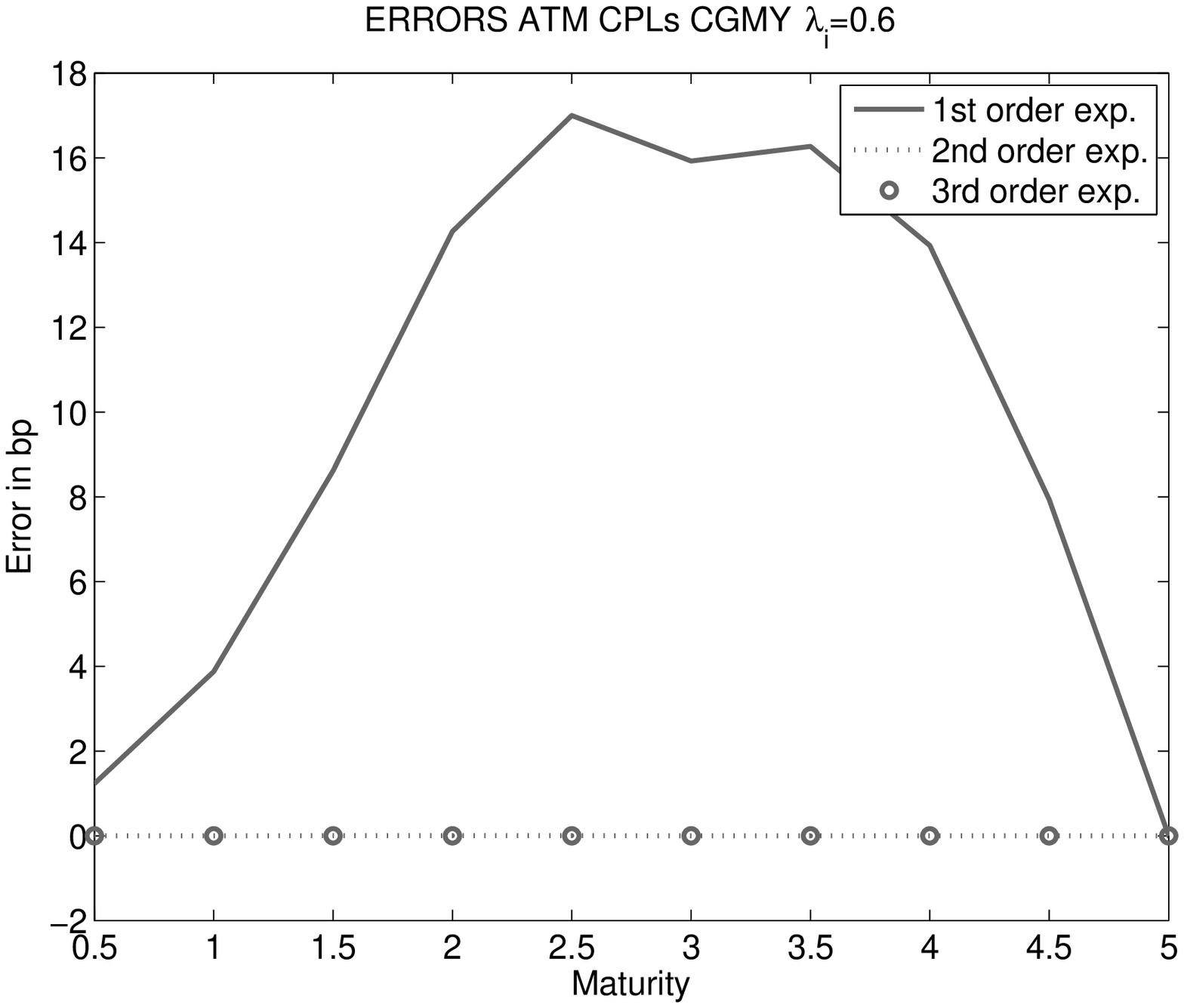}
\includegraphics[width=6.25cm]{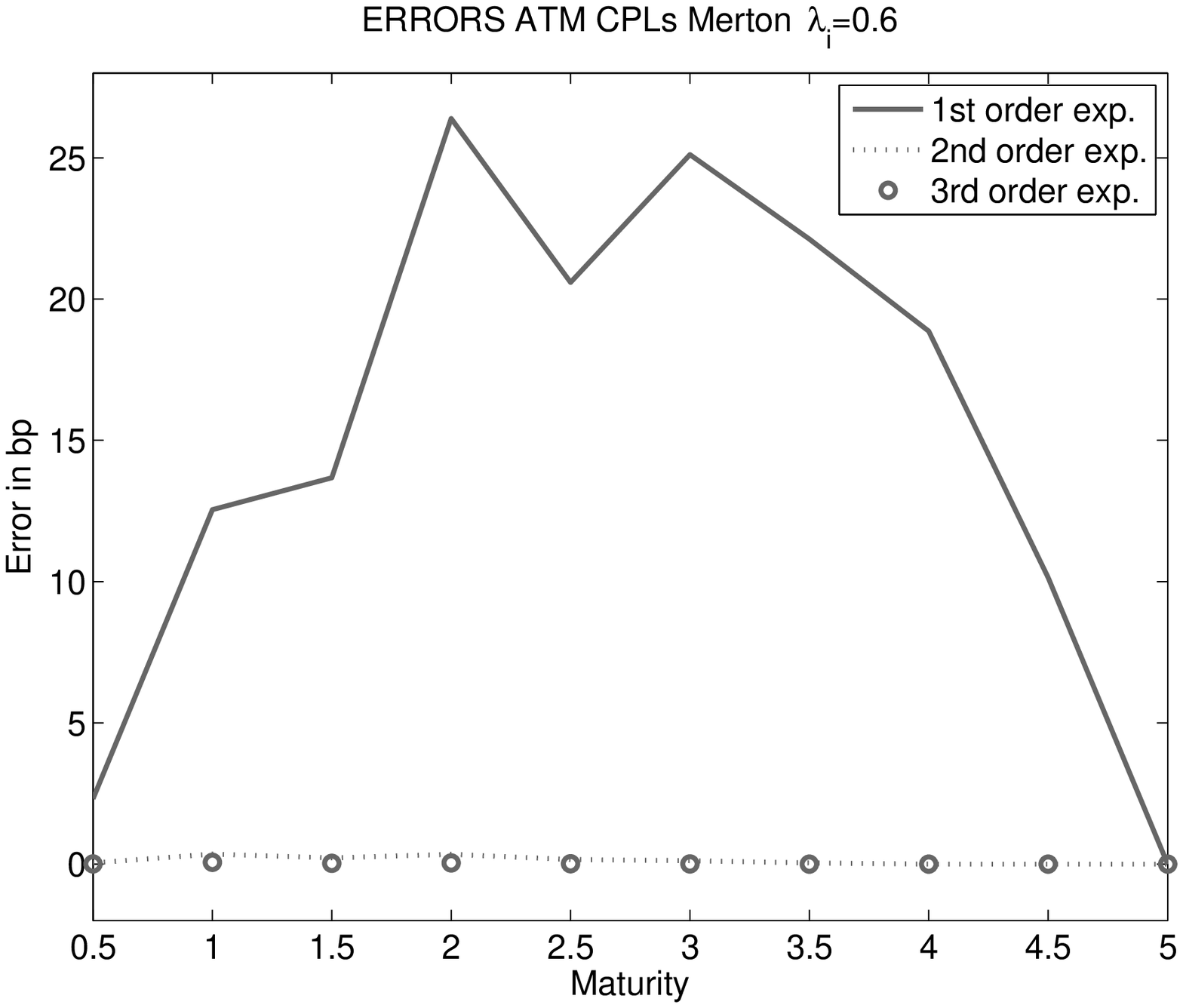}\\
 \caption{Drift expansion: high volatility scenario.}
 \label{fig:Drift-High}
\end{figure}

Due to the complexity of calculating the true drift we limit ourselves to
setting $N=10$, corresponding to a 5 year term structure. Furthermore we
consider volatility structures constant and flat at $\lambda_i=0.2$ and
$\lambda_i=0.6$ respectively. We simulate 10000 paths and plot the absolute
difference between the prices from the drift expansions and the price without
expansion (i.e. the full drift in \eqref{dr}) in Figures \ref{fig:Drift-Low}
and \ref{fig:Drift-High}. Each Monte Carlo simulation is done using the same
random shocks for each method, thus eliminating the Monte Carlo noise as an
error source. The figures demonstrate that the effect of the truncation depends
mostly on the level of volatility $\lambda_i$ and less in the choice of product
to price or the driving process.
Furthermore, we notice that for low volatility even the first order expansion
can be considered adequate, since the maximum of the absolute error is smaller
than 0.2 bp. Conversely, for the high volatility case, the second order
expansion is necessary to get proper accuracy. However, going to the third
order expansion or beyond appears to be unnecessary as there is no visible gain
in accuracy ($<10^{-5}$ bp). Hence, in the next sections we will use the second
order drift expansion as our benchmark case since any resulting error is small
enough to be disregarded.

In Table \ref{table:CPUTIMES}, CPU times are shown when simulating 10000 paths
on an Intel i7 PC running Matlab. Here we can see that highly significant
speed-up is achieved when truncating the higher order drift terms, whereas the
decrease in speed when taking higher order approximations into account is
relatively negligible. The CGMY is slower than the Merton model due to the much
higher jump intensity needed in its approximation. We conjecture that the
efficiency can be improved using the methods of \cite{KohatsuHigaTankov2010},
but this lies outside the focus of this article.
\begin{table}[ht!]
\label{table:CPUTIMES}
\begin{tabular}{ccccc}
& Full Drift & 1st order & 2nd order & 3rd order\\
\hline
Merton & 358.5 & 3.95 & 4.48 & 4.79\\
CGMY & 471.9 & 16.29 & 16.59 & 16.74\\
\hline
\end{tabular}~\\[1ex]
\caption{CPU Times (secs) for 10000 paths}
\end{table}

Finally, to conclude the subsection we should also mention that pricing errors
for swaptions and ratchet caplets(not shown here) are of similar order of
magnitude as in case of
caplets.

\subsection{Performance of the log-\lev approximations}

\begin{figure}[ht!]
 \centering
\includegraphics[width=6.25cm]{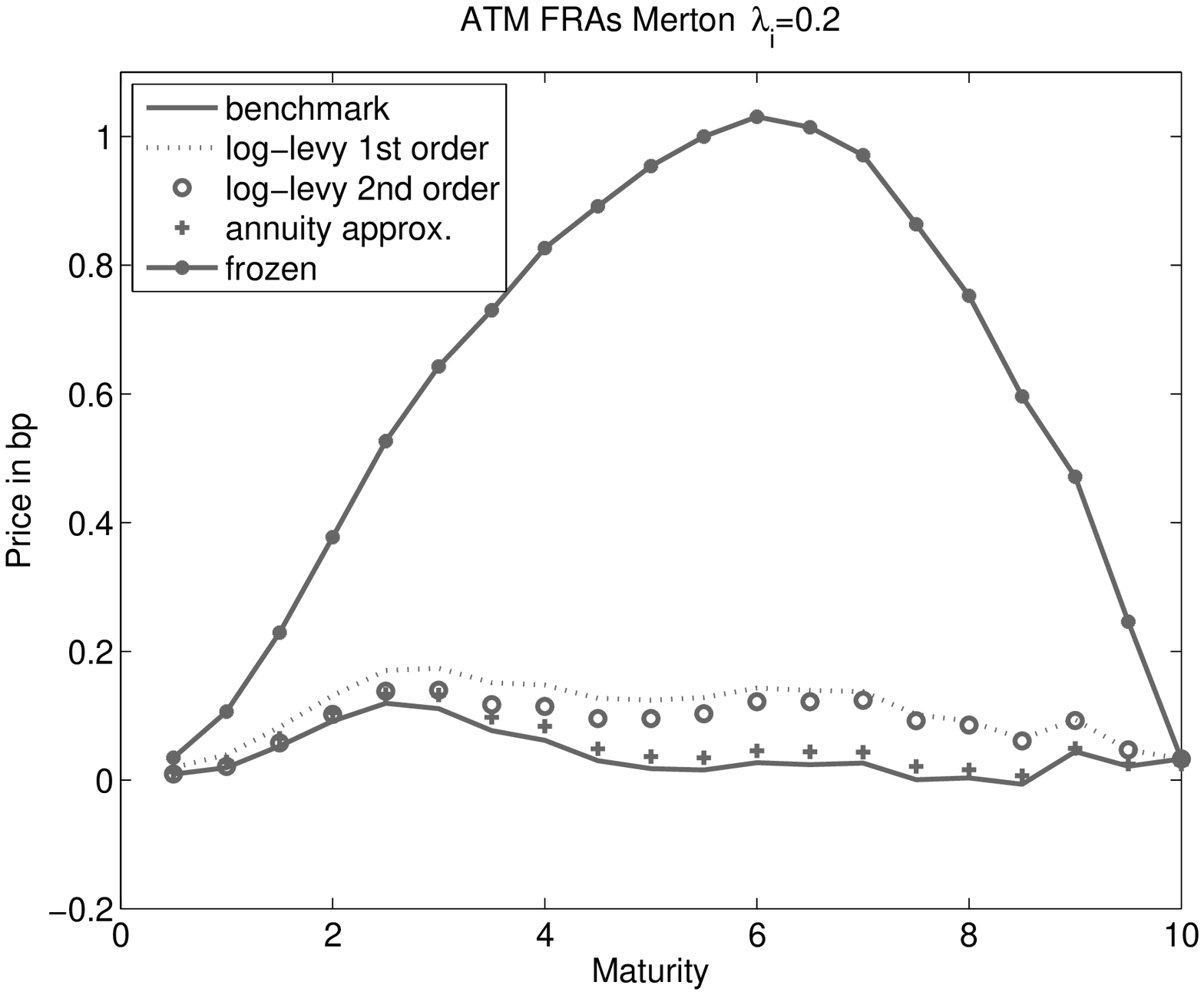}
\includegraphics[width=6.25cm]{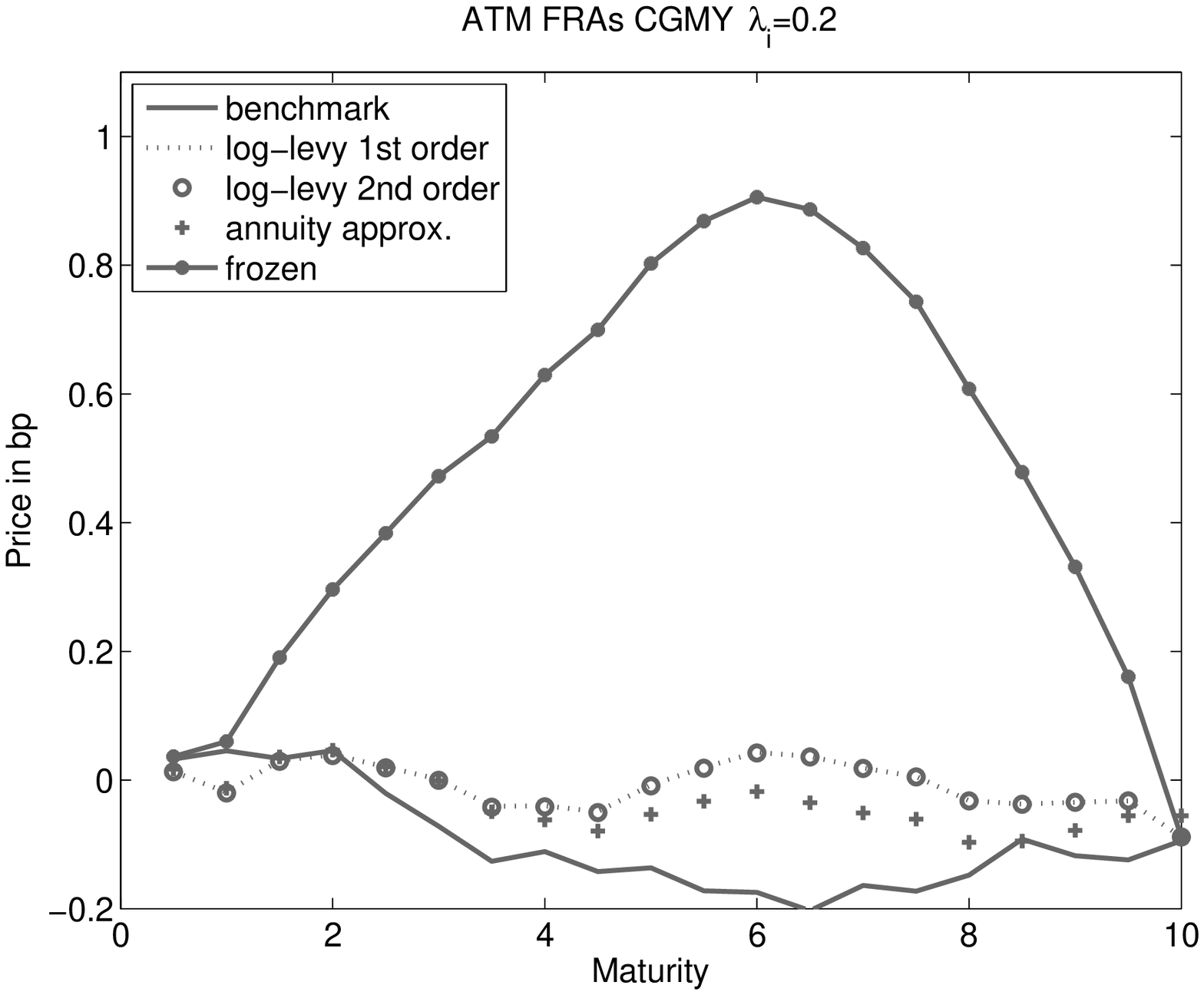}\\
\includegraphics[width=6.25cm]{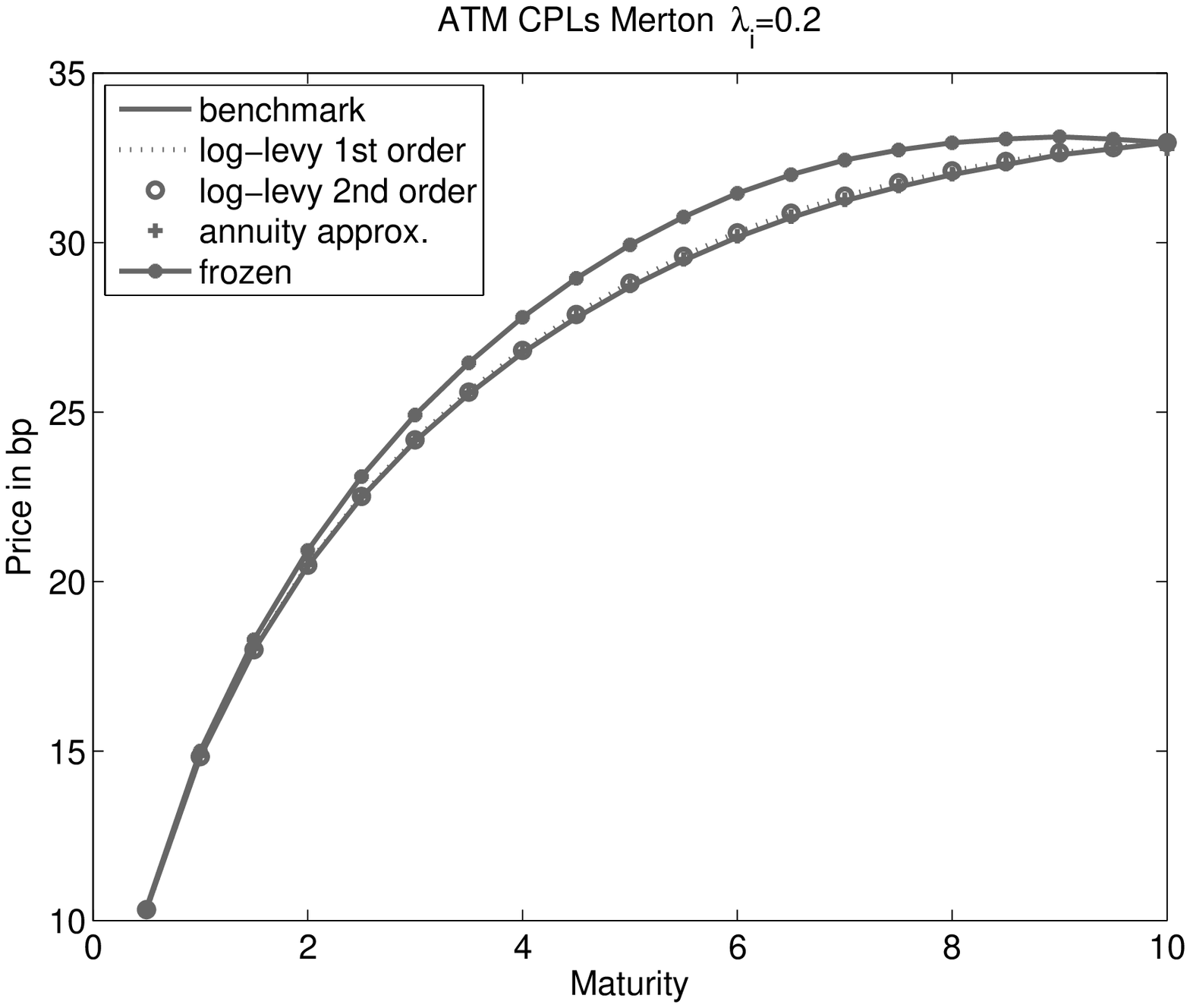}
\includegraphics[width=6.25cm]{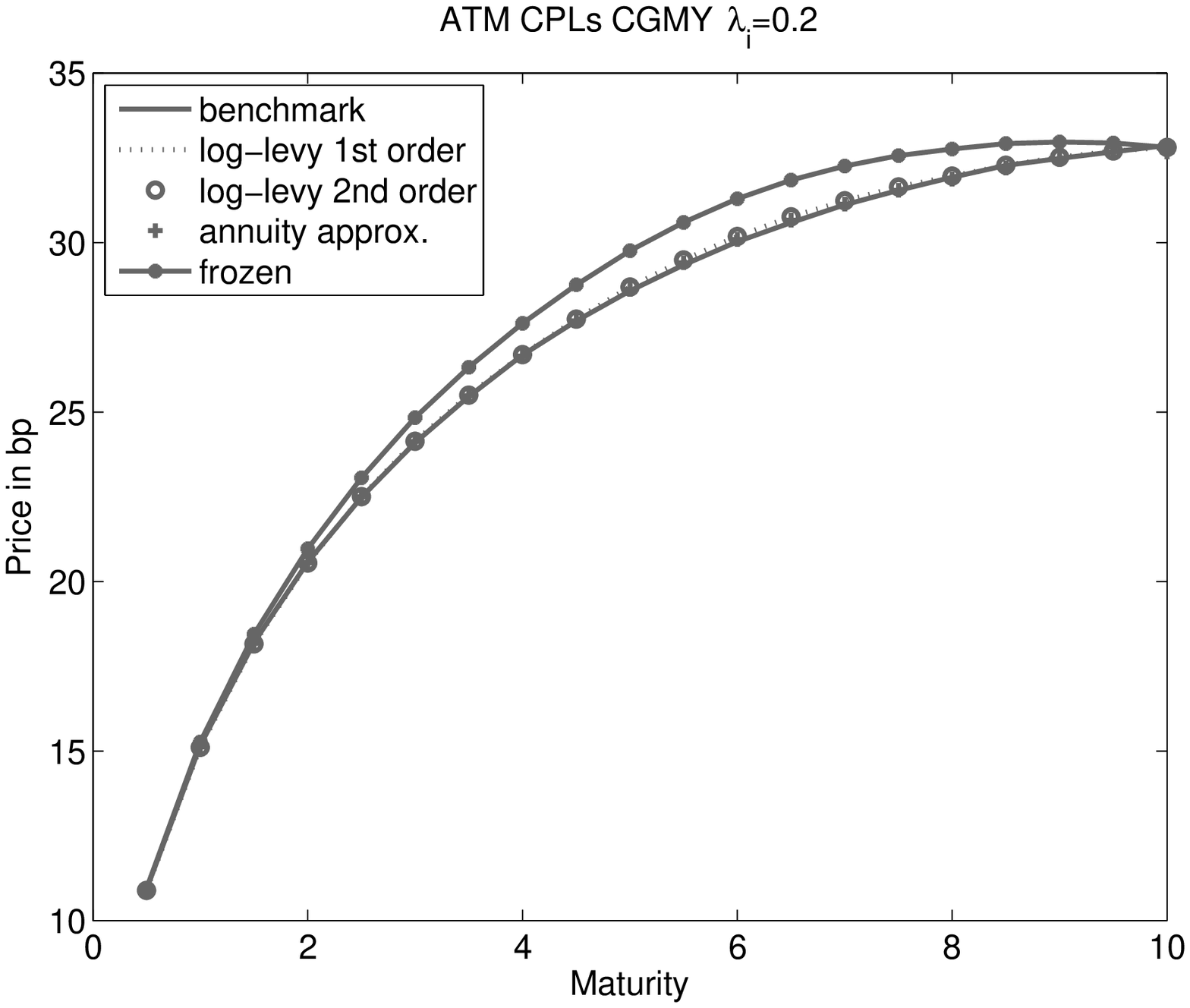}\\
 \caption{Prices for the Merton and CGMY models.}
 \label{fig:Log-Levy-Low-a}
\end{figure}

\begin{figure}[ht!]
 \centering
\includegraphics[width=6.25cm]{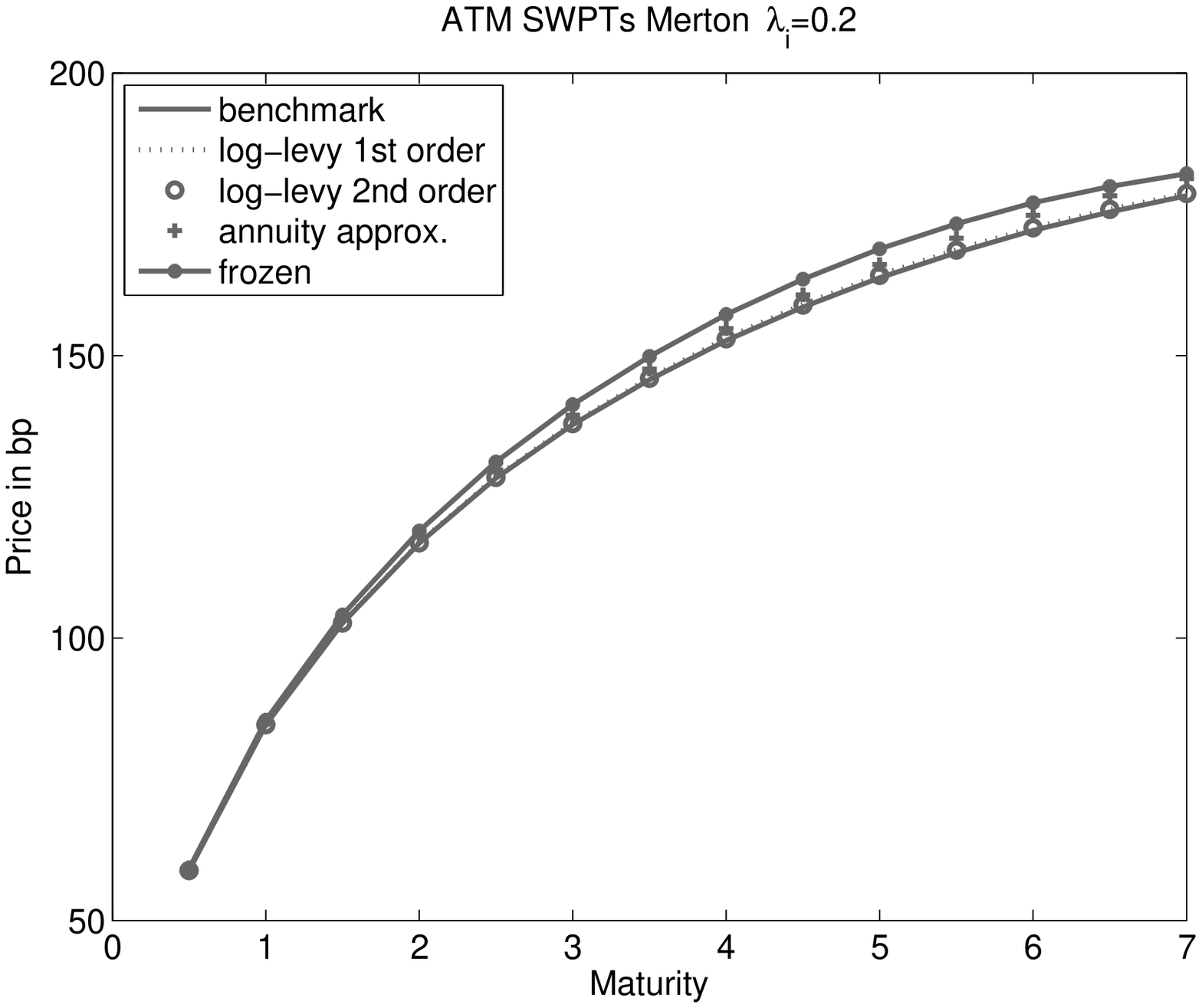}
\includegraphics[width=6.25cm]{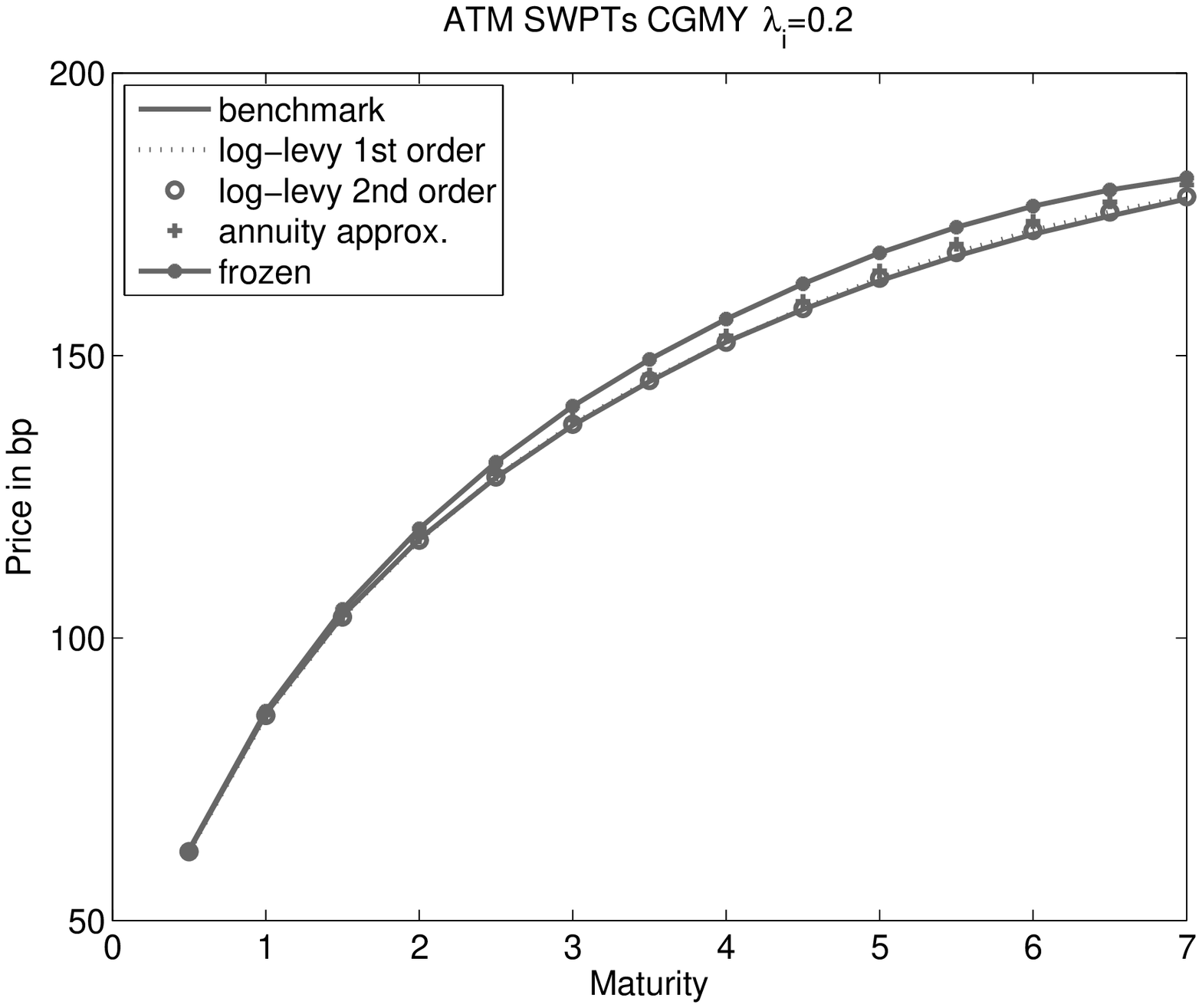}\\
\includegraphics[width=6.25cm]{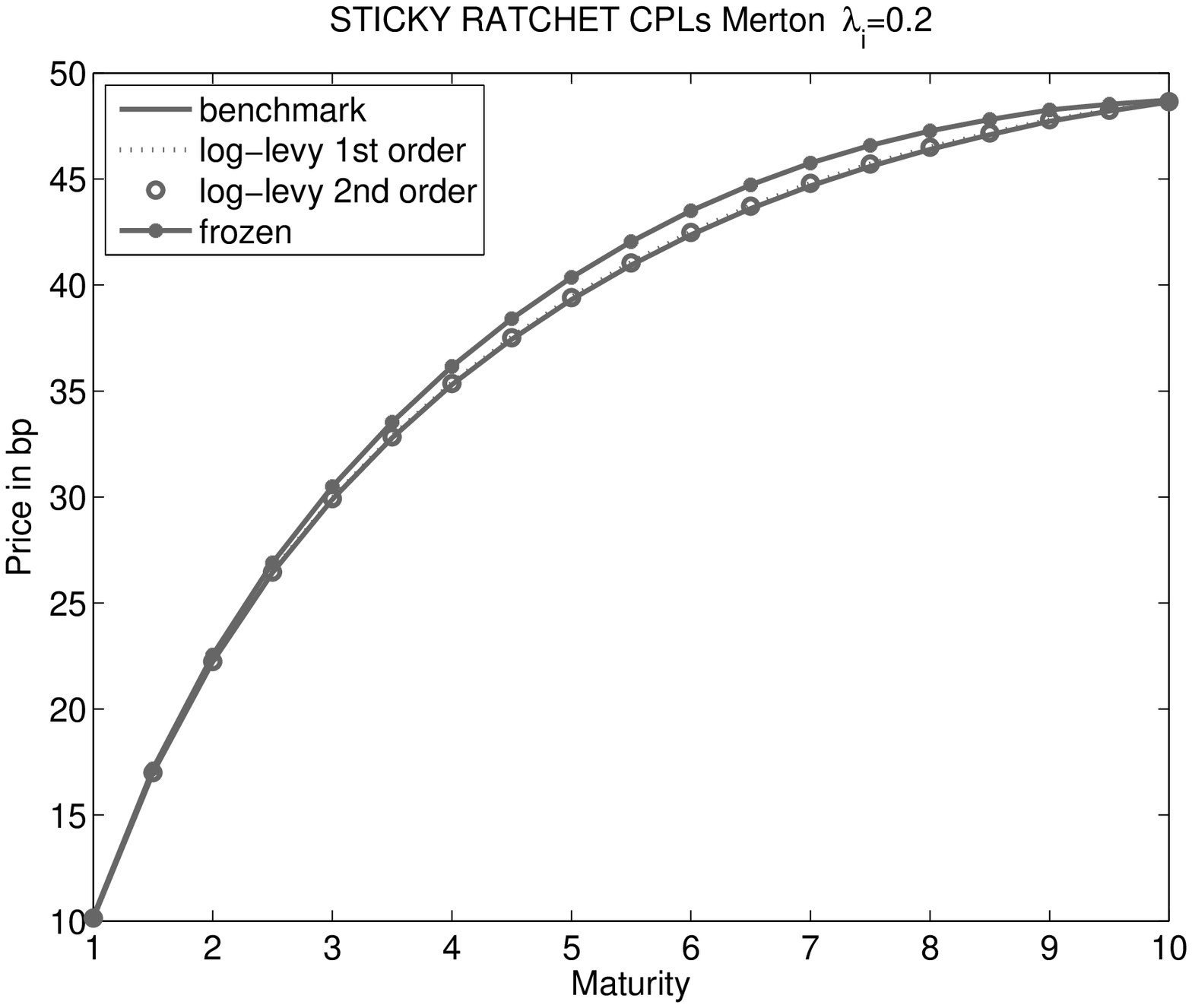}
\includegraphics[width=6.25cm]{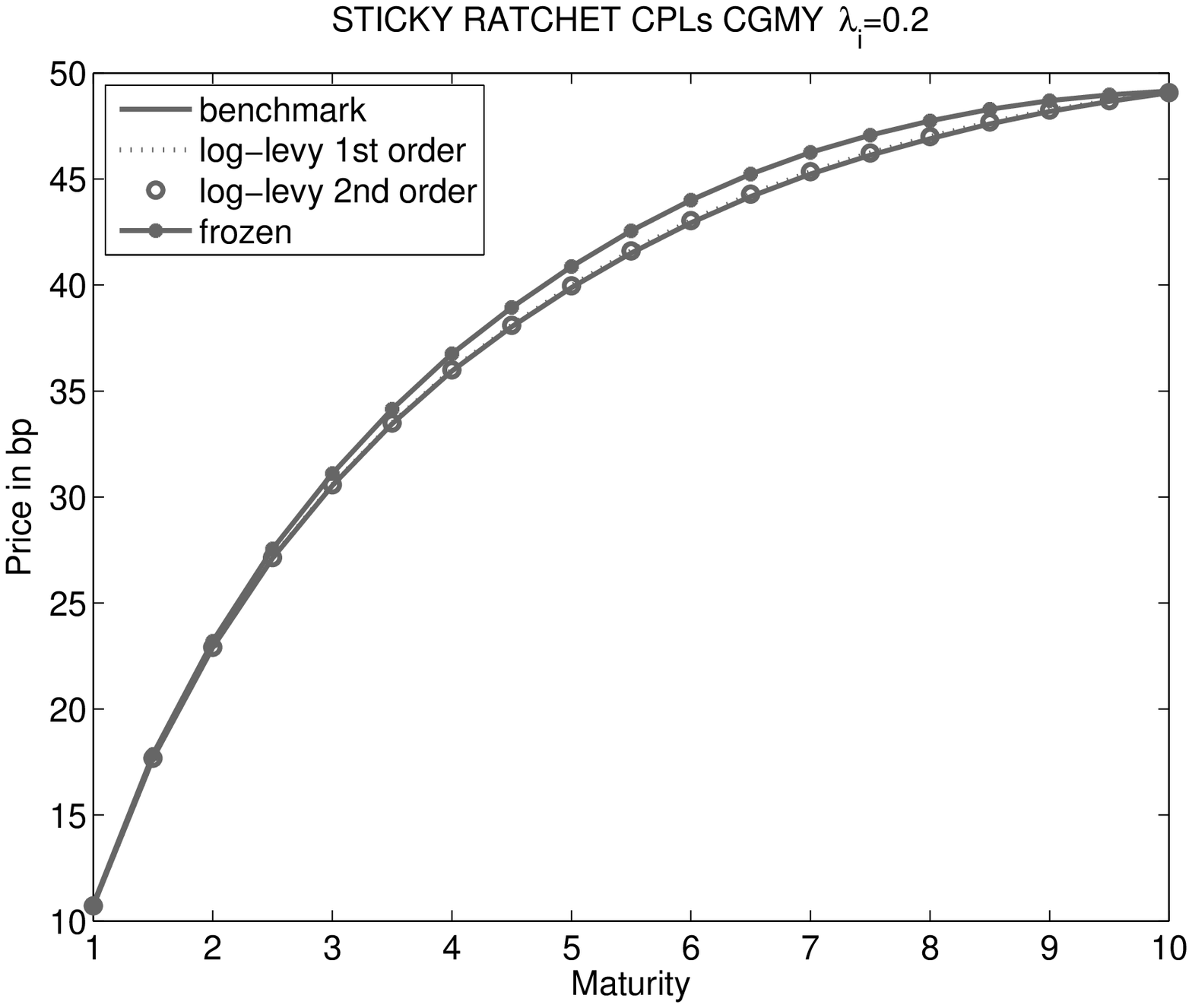}
 \caption{Prices for the Merton and CGMY models.}
 \label{fig:Log-Levy-Low-b}
\end{figure}

Next we study the performance of the log-\lev approximations. We increase the
number of rates to the more realistic setting of $N=20$ and consider the pricing
of FRAs, caplets, sticky ratchet caplets and swaptions. We consider swaptions on
swap rates over the periods $(T_i,T_i + 3)$ years. Since we have established
that errors from the drift expansion can be disregarded, we consider as the
benchmark case the second order drift expansion studied in the previous section.
In Figures \ref{fig:Log-Levy-Low-a} and \ref{fig:Log-Levy-Low-b} we plot prices
from the frozen drift, the first and second order log-\lev approximations of
section \ref{LogLevy}, and include the annuity approximation of the following
section for completeness (for the path-independent derivatives). We use both the
Merton and the CGMY model. We can observe that the frozen drift is consistently
beaten by both the 1st and 2nd order approximation in both models and for all
four products. The 1st and 2nd order log-\lev approximations have a quite
similar performance suggesting that second order approximation may not be
necessary. Note that other parameter values (higher/lower intensity for Merton
and fatter tails/slower tail decay for CGMY) have also been studied and again
the results are qualitatively the same.

Concluding, the log-\lev approximations offer an alternative to the Euler (or
predictor-corrector) discretization of the actual dynamics which can be
simulated faster and yields almost as accurate options prices.

\section{Approximation of annuities}
\label{annuity-section}

In the lognormal LIBOR market model, it is well documented that problems may
occur for high volatilities due to a proportionally large Monte Carlo variance
in the annuity term used for discounting under the terminal measure, see
\cite{Beveridge10} and \cite{GatarekBachertMaksymiuk06}. Motivated
by this numerical problem, we will derive an approximation of the annuity term
in the spirit of \cite[\S 10.13]{GatarekBachertMaksymiuk06}.

Let us define the annuity term
\begin{align}\label{annuity}
A_i(t) = \prod_{j=i+1}^N (1+\delta_j L_j(t)),
\end{align}
and consider the vector of log-\lib rates $G=[G_{i+1},\dots,G_N]$. We define a
function $f:\R^{N-i}\to\R$ such that
\begin{align*}
(x_{i+1},\dots,x_N)=x \longmapsto \prod_{j=i+1}^N (1+\delta_j \e^{x_j}).
\end{align*}
The partial derivatives of $f$ are provided by
\begin{align*}
f_k(x) = \frac{\partial}{\partial x_k}f(x)
       = \prod_{\substack{j=i+1\\j\ne k}}^N (1+\delta_j \e^{x_j}) \delta_k
\e^{x_k}
       = f(x)\frac{\delta_k \e^{x_k}}{1+\delta_k \e^{x_k}},
\end{align*}
for all $i+1\le k\le N$, while we obviously have that
\begin{align}
f(G(t)) = A_i(t).
\end{align}

Applying It\^o's formula to $f(G)$, we have that
\begin{align}\label{f-ann}
f(G(t))
  = A_i(t)
 &= A_i(0) + \sum_{j=i+1}^N \int_0^t f_j(G(s-)) \ud G_j(s) \nonumber\\
 &\quad + \frac12 \sum_{j,k=i+1}^N \int_0^t f_{j,k}(G(s-))
           \ud \langle G_k,G_j(s)\rangle^c(s) \nonumber \\
 &\quad + \sum_{s\le t} \left\{ \Delta f(G(s))
           - \sum_{j=i+1}^N f_j(G(s-)) \Delta G_j(s)\right\}.
\end{align}
Noting that the annuity is a $\P_*$-martingale, we will focus on the martingale
parts of \eqref{f-ann} in the sequel. Using \eqref{intL} and the fact that $H$
is also a $\P_*$-martingale, we get that the \textit{martingale} part of the
first summand is
\begin{align*}
\sum_{j=i+1}^N \int_0^t f_j(G(s-)) \lambda_j(s) \ud H(s)
 &= \sum_{j=i+1}^N \int_0^t \frac{\delta_j L_j(s-)}{1+\delta_j L_j(s-)}
     f(G(s-)) \lambda_j(s) \ud H(s) \nonumber\\
 &= \int_0^t A_i(s-)
    \sum_{j=i+1}^N \frac{\delta_j L_j(s-)}{1+\delta_j L_j(s-)}\lambda_j(s) \ud
H(s).
\end{align*}
The second summand is omitted, while the final summands yields that
\begin{align}
& \sum_{s\le t} \left\{ \Delta f(G(s))
           - \sum_{j=i+1}^N f_j(G(s-)) \Delta G_j(s)\right\} \nonumber\\
&= \sum_{s\le t} \left\{ \Delta A_i(s)
           - A_i(s-) \sum_{j=i+1}^N  \frac{\delta_j L_j(s-)}{1+\delta_j L_j(s-)}
             \Delta G_j(s)\right\} \nonumber\\
&= \sum_{s\le t} \left\{ \Delta A_i(s)
           - A_i(s-) \sum_{j=i+1}^N  \frac{\delta_j L_j(s-)}{1+\delta_j L_j(s-)}
             \lambda_j(s)\Delta H(s)\right\} \nonumber\\
&= \int_0^t\int_{\R^m} \left\{ A_i(s) - A_i(s-)
           - A_i(s-) \sum_{j=i+1}^N  \frac{\delta_j L_j(s-)}{1+\delta_j L_j(s-)}
             \lambda_j(s)x\right\} \mt\dsdx \nonumber\\
&\,\, -\int_0^t\int_{\R^m} \left\{ A_i(s) - A_i(s-)
           - A_i(s-) \sum_{j=i+1}^N  \frac{\delta_j L_j(s-)}{1+\delta_j L_j(s-)}
             \lambda_j(s)x\right\} F(s,\dx)\ds,
\end{align}
where the quantity $A_i(s)$ in the last two integrals should be understood as
\begin{align}\label{annuity-jm}
A_i(s) = \prod_{j=i+1}^N
      \left( 1+\delta_j\exp\big\{G_j(s-)+\lambda_j^{\mathsf{T}}(s)x\big\}
\right).
\end{align}

Collecting all the pieces together, we have that the annuity $A_i$ satisfies the
following integrated SDE
\begin{align}\label{annuity-SIE}
A_i(t)
 &= A_i(0) + \int_0^t A_i(s-) \Lambda_i(s-) \ud H(s) \nonumber\\
 &\quad + \int_0^t\int_{\R^m} \left\{ A_i(s)-A_i(s-)
           - A_i(s-) \Lambda_i(s-)\right\} \mt\dsdx \nonumber\\
\intertext{or, equivalently}
A_i(t)
 &= A_i(0) + \int_0^t A_i(s-) \Lambda_i(s-) \ud H(s) \nonumber\\
 &\quad + \int_0^t\int_{\R^m} A_i(s-)\left\{ \frac{A_i(s)}{A_i(s-)} - 1
           - \Lambda_i(s-)x\right\} \mt\dsdx,
\end{align}
where
\begin{align}
\Lambda_i(s-)
 = \sum_{j=i+1}^N  \frac{\delta_j L_j(s-)}{1+\delta_j L_j(s-)}
             \lambda_j(s).
\end{align}
The solution of the SDE \eqref{annuity-SIE} is the stochastic exponential, thus
we get that
\begin{align}\label{annuity-exp}
A_i(t)
 &= A_i(0) \exp \left( \int_0^t \Lambda_i(s-) \ud W(s)
          - \frac12 \int_0^t \Lambda_i^{\mathsf{T}}\Lambda_i(s-) \ds \right.
\nonumber\\
 &\qquad\qquad\qquad +
   \int_0^t \int_{\R^m} \left\{ \frac{A_i(s)}{A_i(s-)} - 1 \right\}\mt\dsdx
  \\\nonumber
 &\qquad\qquad\qquad \left.
  - \int_0^t \int_{\R^m}\left( \log\left\{ \frac{A_i(s)}{A_i(s-)} \right\}
          - \frac{A_i(s)}{A_i(s-)} +1 \right)\mu\dsdx\right),
\end{align}
where again $A_i(s)$ should be understood as in \eqref{annuity-jm}. By freezing
the random terms in the drifts and jump sizes in the above dynamics we get an
alternative approximation for the annuity term. Note that the resulting
approximation is also a \textit{log-L\'evy approximation}.

We can now use this approximation to price caplets and swaptions, noting that
their respective payoffs can be written in terms of annuities:
\begin{align}
\mathbb{C}_0
 &= B_{N+1}(0)\,
    \E_{\P_*}\left[ \frac{A_{i}(T_{i+1})}{A_{i}(T_{i})} \Big( A_{i-1}(T_{i}) 
                   - \left(1+\delta_{i}K\right) A_{i}(T_{i})\Big)^{+}\right],\\
\mathbb{S}_{0}
 &= B_{N+1}(0)\, \E_{\P_*} \left[\left( - \sum^m_{k=i}
    c_k A_{k-1}(T_i)\right)^+\right],
\end{align}
where the $c_k$'s are defined in \eqref{ck}. A similar expression can be derived
for the sticky ratchet caplet.

\subsection{Performance of the annuity approximation}
\begin{figure}[ht!]
 \centering
\includegraphics[width=6.25cm]{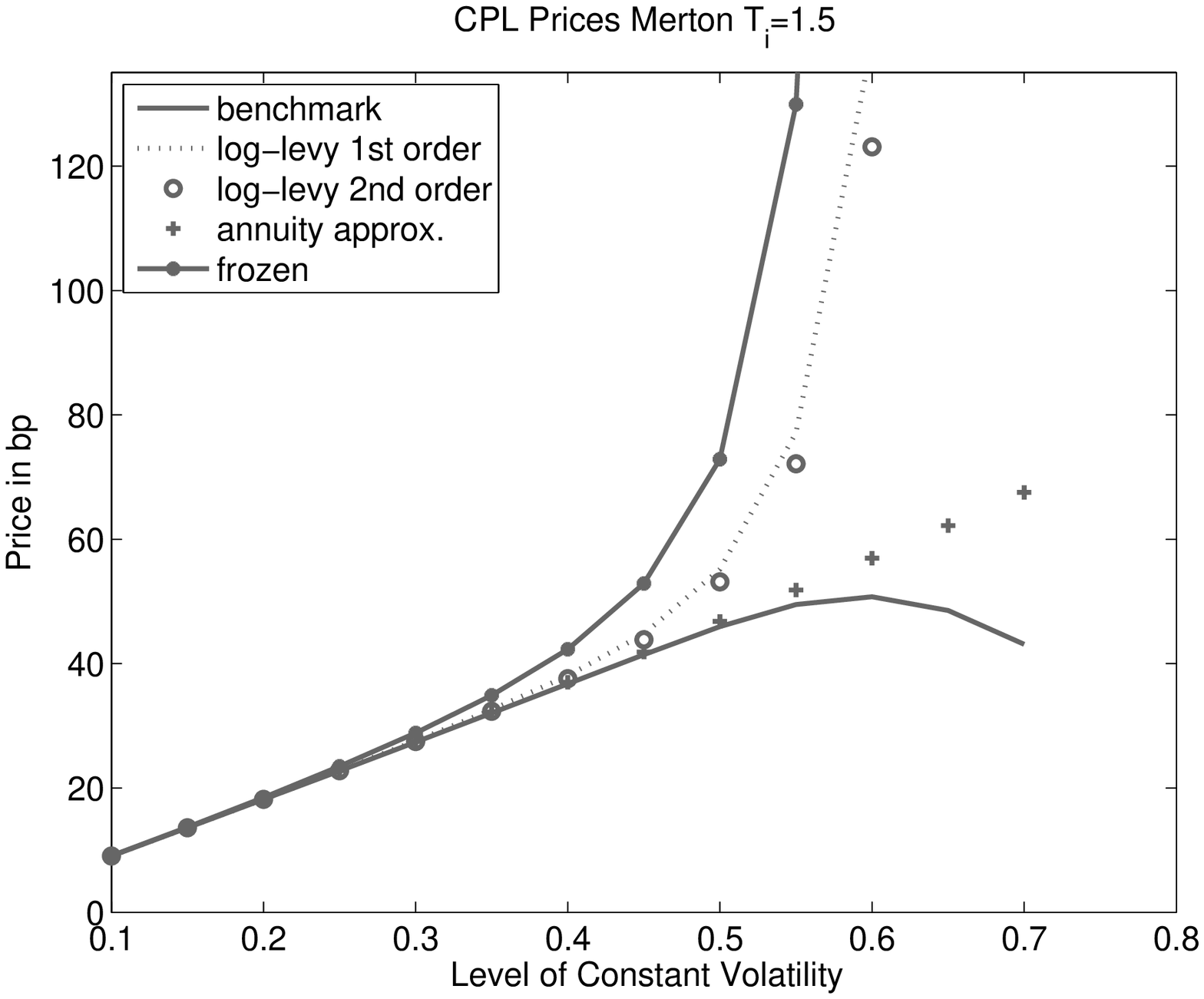}
\includegraphics[width=6.25cm]{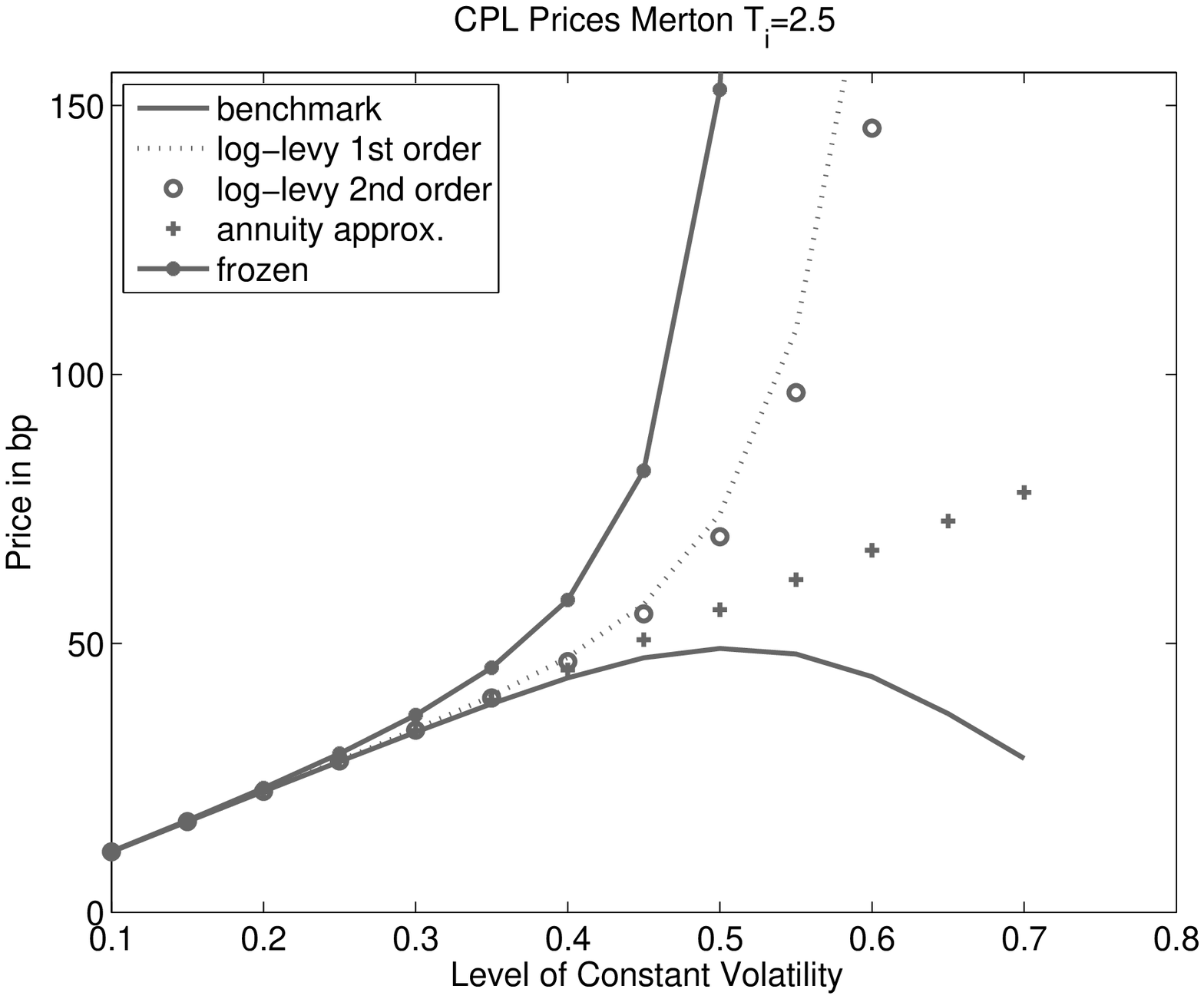}\\
\includegraphics[width=6.25cm]{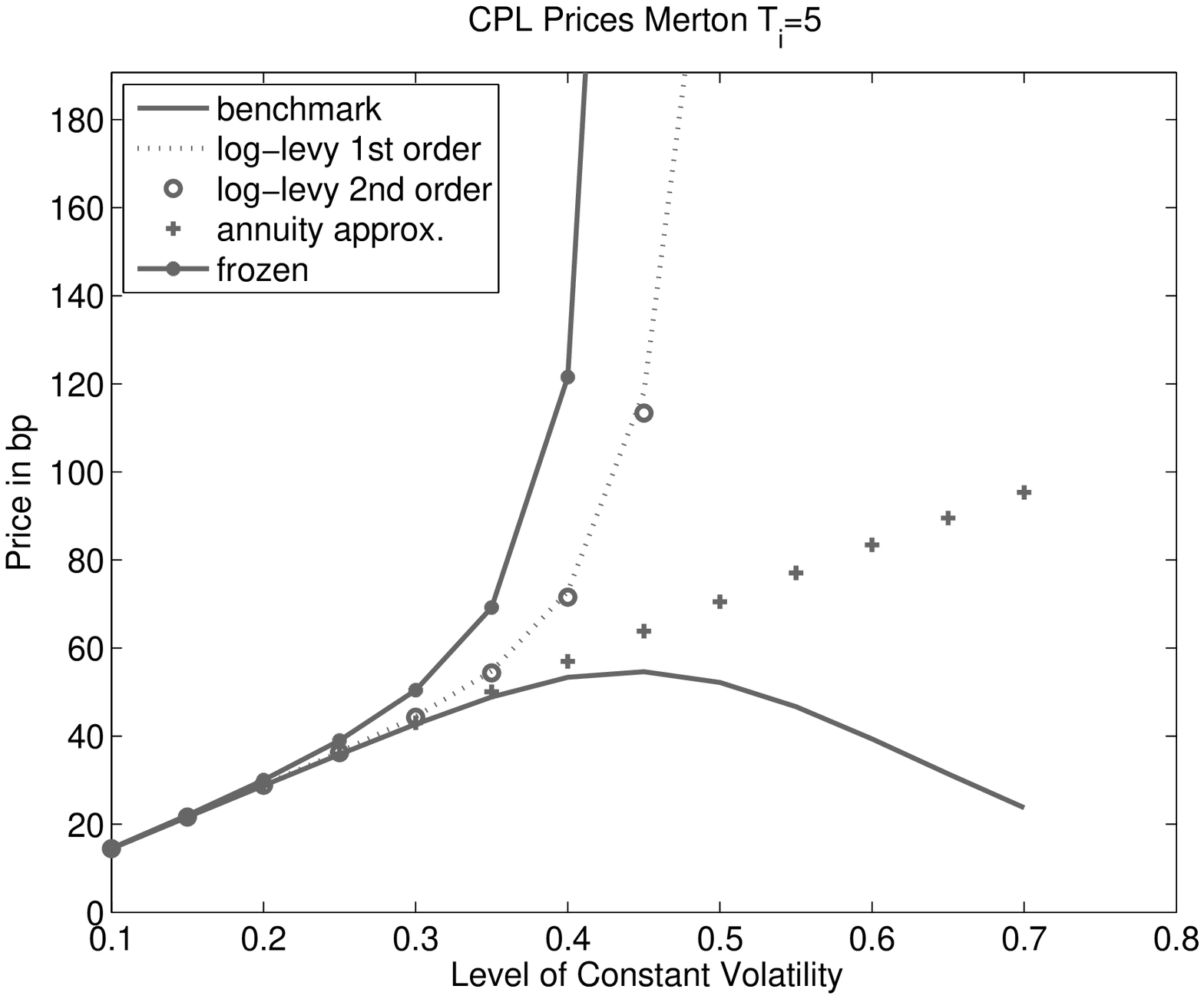}
\includegraphics[width=6.25cm]{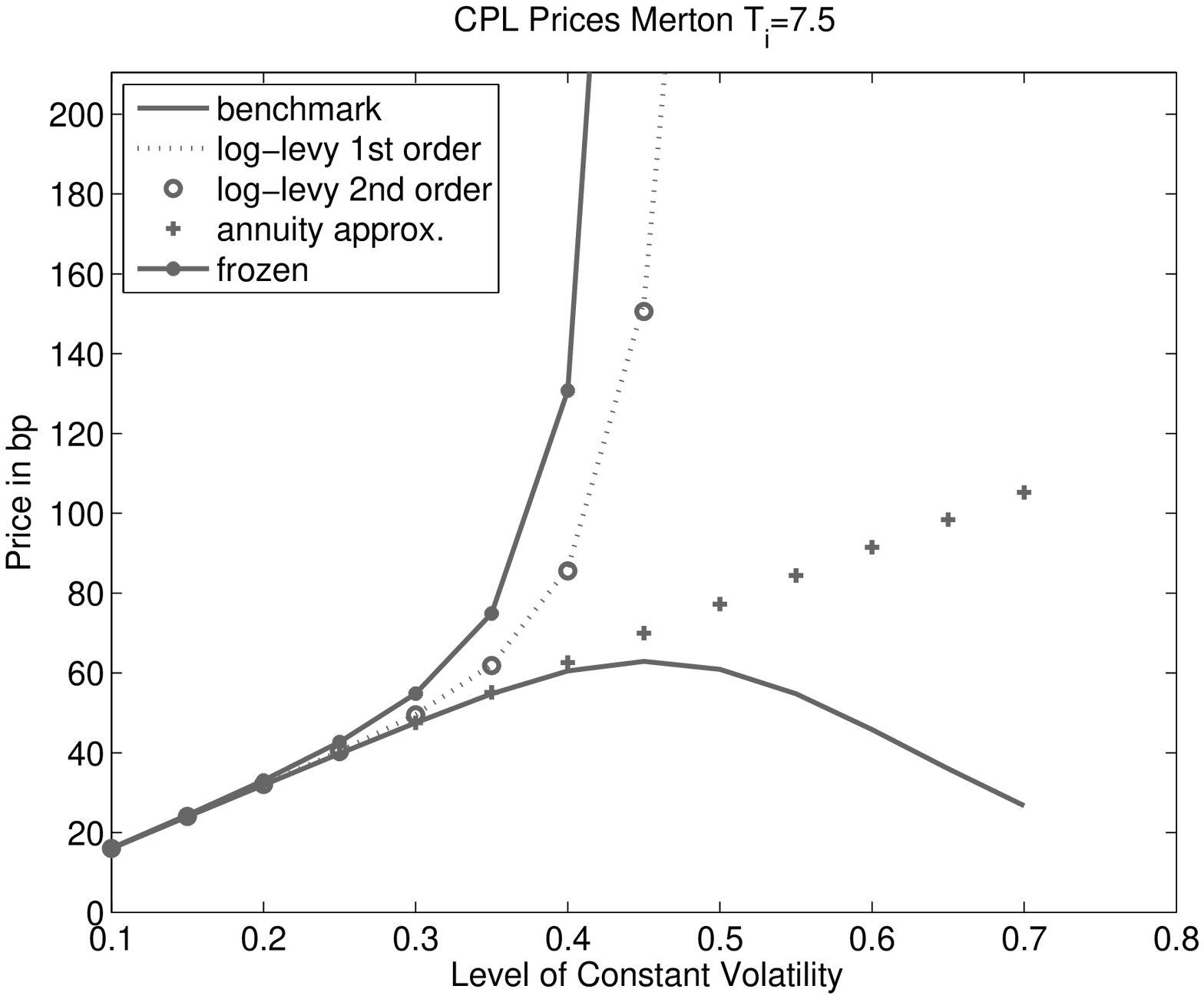}\\
 \caption{Caplet prices as a function of volatility ($N=20$).}
 \label{fig:volgraph-merton}
\end{figure}
In Figure \ref{fig:volgraph-merton}, the quality of the various approximations
is studied for a number of at-the-money caplets as a function of the volatility.
As before we set the number of rates to $N=20$, and simulate 50000 paths for
each volatility level. The plot is for the Merton model while the results are
similar for CGMY. Using that at-the-money call option prices are increasing and
roughly linear functions of volatility (see for example \cite{Wilmott1998},
\cite{BrennerSubrahmanyam1994} and \cite{BackusForesiWu2004} for the case of
non-Gaussian distributions), we can observe that only the annuity approximation
produces sensible option prices at all levels of volatility. Moreover, even the
benchmark case fails when volatility grows beyond $30\%$, meaning that the Monte
Carlo simulation has failed to converge.  The frozen drift fails at even lower
levels of volatility, while the log-\lev approximations fail at a higher level,
similar to the benchmark case. The annuity approximation works for all (higher)
levels and also, as we have seen in Figures \ref{fig:Log-Levy-Low-a} and
\ref{fig:Log-Levy-Low-b}, for the low
levels.  One should therefore be careful when the average (across maturity)
at-the-money implied volatilities are above 30\% which is indeed the case in the
current market for USD denominated LIBOR caplets where volatilities range from
roughly 80\% in the short end to 25\% in the long end (source: Bloomberg). 

\begin{figure}
 \centering
 \includegraphics[width=8.25cm]{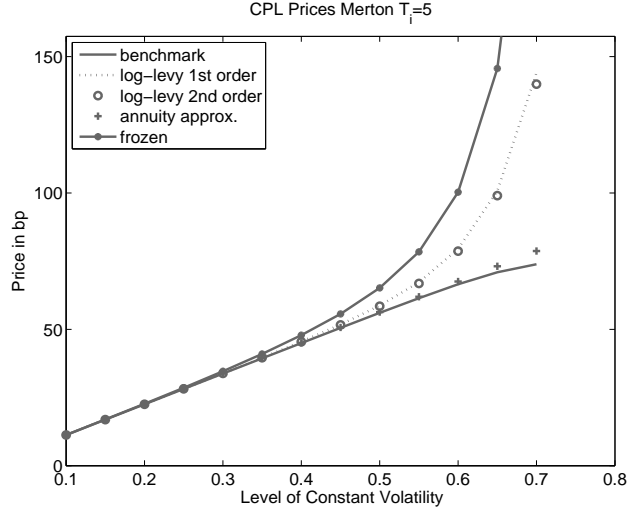}\\
 \caption{Caplet prices as a function of volatility ($N=10$).}
 \label{fig:volgraph-mertonN10}
\end{figure}

Moreover, in Figure \ref{fig:volgraph-mertonN10} we observe that this problem
becomes significantly less severe when limiting the number of rates to 10 with
$\delta_i=1$ instead of 20 with $\delta_i=0.5$. Needless to say, limiting the
number of rates is rarely a possibility in practice.

In order to intuitively understand why this approximation performs better in the
high volatility case than the other methods (e.g. the standard Euler scheme or
the log-\lev approximations), let us just concentrate on the lognormal case. We
have from \eqref{annuity-exp} that
\begin{align}\label{annuities-var-I}
\log A_i(t) \cong \sum_{j=i+1}^N  \frac{\delta_j L_j(0)}{1+\delta_j L_j(0)}
                  \lambda_j \cdot \sqrt{t}\,\mathcal{N}
                + \text{deterministic terms},
\end{align}
where $\mathcal{N}$ denotes a standard normal random variate. On the other hand,
from \eqref{annuity}, we get that
\begin{align}\label{annuities-var-II}
\log A_i(t) \cong \sum_{j=i+1}^N  \delta_j
                  \exp\left( \lambda_j \cdot \sqrt{t}\,\mathcal{N}
                             + \text{random terms} \right),
\end{align}
where actually the method of approximation will only affect the random terms. We
can easily conclude from \eqref{annuities-var-I} and \eqref{annuities-var-II}
that the variance of the annuity approximation is significantly lower that the
variance of the standard representation, which results in the faster convergence
of the Monte Carlo method. Thus, the annuity log-\lev approximation should be
interpreted as a \textit{variance reduction} technique for the \lib market
model.

\section{Economically meaningful multi-dimensional \lev measures via
subordination} \label{multi-dim}

Next, we reflect on the properties the driving process should have for practical
applications and provide some recommendations. In an economically realistic \lev
\lib model the very structure of the \lev measure is important. Since, from an
economic point of view, any jump in the daily rate typically affects all
segments of the yield curve, we require in our modeling that, \textit{at a jump
time, all the LIBORs jump, not only the first or second half of the \lib curve
for example.} Moreover, this requirement should be fulfilled regardless of the
structure of the loading factors $\lambda_{i}$; the latter may be inferred from
some calibration procedure for instance. A natural way to meet this property is
to take \lev measures which are \textit{absolutely continuous}. In a
jump-diffusion setting this can be easily established by taking as \lev measure
the product of one dimensional absolutely continuous probability measures
$p_{i}$, i.e. 
\begin{equation}\label{pro}
\nu(\dx)=p_{1}(\dx_{1})\cdot\cdot\cdot p_{m}(\dx_{m});
\end{equation}
see \cite{BelomestnySchoenmakers06}. In this paper we consider \lib models
based on \lev processes with possibly infinite activity, thus having available
flexible and realistic \lib models possibly without Wiener part (i.e.
$\alpha\equiv0$).
However, \lev measures of infinite activity cannot be obtained by simply taking
the product of a set of one-dimensional \lev measures of infinite activity.
Nonetheless, we seek for absolutely continuous infinite activity \lev measures
such that the entailed jump processes  maintain certain (weak) independence
properties. Such measures may be constructed by \textit{Brownian subordination}
(see e.g.  \cite{ContTankov03}) as outlined below.

Let $W$ be a Wiener process on $\mathbb{R}^{m}$. The characteristic function of
$W(t)$ is given by
\[
\E\left[ \e^{izW(t)} \right]
 = \e^{-\frac{t}{2}\left\Vert z\right\Vert^{2}}
 =: \e^{t\Psi(z)},\quad z\in\R^m.
\]
We now consider a subordinator $(S_{t})_{t\geq0}$ on $\mathbb{R}_{+},$ with \lev
triplet $(0,0,\rho),$ and with Laplace exponent $\Xi$, i.e.
\[
\E\left[ \e^{uS_{t}} \right]
 = \e^{t \Xi(u)}
:=\exp\Big(t\int_{\left(0,\infty\right)}
   \left( \e^{su}-1 \right) \rho(\ds)\Big),\quad u\leq0.
\]
Then the $m$-dimensional process $Y$ defined by
\[
Y(t):=W(S_{t})%
\]
has characteristic function%
\begin{align*}
\E\left[ \e^{iz^{\mathsf{T}}Y(t)}\right]
 &= \E\left[\E\left[ \e^{iz^{\mathsf{T}}W(S_{t})}|S_{t}\right]\right]
  = \E\left[\e^{S_{t}\Psi(z)}\right]
  =\e^{t\Xi(\Psi(z))}\\
 &=\exp\left[ t\int_{\left(0,\infty\right)}
    \left(\e^{s\Psi(z)}-1\right) \rho(\ds)\right] \\
 &=\exp\left[t\int_{\left(0,\infty\right)}
    \left(\e^{-\frac{s}{2}\left\Vert z\right\Vert^{2}}-1\right)\rho(\ds)\right]
 =:\exp\left[t\Phi(z)\right]
\end{align*}
As a result, $Y$ is a pure jump martingale \lev process with \lev measure
$\nu^{Y}$ satisfying
\begin{equation}\label{cum}
\Phi(z)
 =\int_{\left(0,\infty\right)}
   \left(\e^{-\frac{s}{2}\left\Vert z\right\Vert^{2}}-1\right)\rho(\ds)
 =\int_{\mathbb{R}^{m}}( \e^{\mathfrak{i}z^{\mathsf{T}}x}
         - 1 -\mathfrak{i}z^{\mathsf{T}}x )\nu^{Y}(\dx).
\end{equation}
It is easily checked that
\begin{equation}\label{subv}
\nu^{Y}(\dx)
 =\int_{0}^{\infty} \frac{1}{\left(\sqrt{2\pi s}\right)^{m}}
   \e^{-\frac{1}{2s}||x||^{2}}\rho(\ds) \dx,
\end{equation}
which is a measure with absolutely continuous support.

\begin{example}
Let $(S_{t})_{t\ge0}$ be the inverse Gaussian subordinator with
\begin{align*}
\rho(\ds)
 = \frac{c\e^{-\lambda s}}{s^{3/2}}1_{\{s>0\}}\ds,
\quad\text{ and }\quad
 \E\left[\e^{uS_{t}}\right]
  =\e^{-2ct\sqrt{\pi}\left(\sqrt{\lambda-u}-\sqrt{\lambda}\right)}.
\end{align*}
Then, \eqref{cum} is known explicitly as
\begin{gather*}
\Phi(z) = -2c \sqrt{\pi}
 \left(\sqrt{\lambda+\frac{\sigma^{2}}{2}\left\Vert z\right\Vert^{2}}
  -\sqrt{\lambda}\right),
\end{gather*}
e.g. see \cite{ContTankov03}.
\end{example}

\begin{example}
Let $(S_{t})_{t\ge0}$ be a L\'evy subordinator with the following properties:
\begin{align*}
\rho(\dt)
 &= C\e^{-\frac{t}{4}G}D_{-Y}(G)1_{\{t>0\}}\dt,\\
\Xi(u)
 &=2C\Gamma(-Y) \left[(G^2-2u)^{Y/2}
   \cos\left(Y\arctan\left(\frac{\sqrt{-2u}}{G}\right)\right)-G^Y\right],
\end{align*}
where $D$ is the parabolic cylinder function. Then, \eqref{cum} is known
explicitly as the L\'evy exponent of the CGMY process, cf. \eqref{kCGMY}, with
$G=M$; see \cite{MadanYor08}.
\end{example}

\begin{remark}
By taking in \eqref{h*} $F(s,\dx):=\nu^Y(\dx)$ with $\nu^Y$ given by
\eqref{subv}, the jump-part of \eqref{h*} is represented by the process $Y$
constructed above. It is easy to see that $Y$ has uncorrelated components,
although they are generally not independent. Indeed, $Y(t)$ has mean zero and we
have that
$$
\E\left[Y^{(k)}(t) Y^{(l)}(t)\right]
 = \E\left[\E\left[W^{(k)}(S_t) W^{(l)}(S_t)\mid S_t\right]\right]=0,
 \quad 1\le k<l\le m.
$$
Thus in contrast to the jump-diffusion situation in
\cite{BelomestnySchoenmakers06} where all components jump at the same time
independently, here the components of $Y$ still jump at the same time but in an
uncorrelated rather than in an independent way.
\end{remark}

\section{Concluding summary}

We have presented a tractable numerical approach to simulate trajectories of a
general L\'evy LIBOR model in an efficient way. By this method we construct
efficient approximations to the computationally demanding drift term in the
L\'evy LIBOR dynamics. We have shown that, due to these these approximations, we
arrive at a significantly more accurate log-L\'evy approximation than the one
obtained by the usual ``frozen drift'' approximation. The performance of the
method is illustrated by several examples. The presentation is embedded in a
flexibly structured multi-factor L\'evy LIBOR model which allows for natural
modeling of mutual LIBOR dependences (via incorporating suitable correlation
structures). As such the paper supports practical implementations of L\'evy
interest rate models that, until now, played mostly an academic role.

\appendix
\section{Computation of the drift}
\label{app-drift}

\subsection{Full expansion in terms of cumulants}
We will derive a representation for the integral term of the drift \eqref{dr}
which does not involve an integration over random terms. Let us denote the
integral term by
\begin{align*}
\mathbb{B}_i
 := \int_{\mathbb{R}^{m}}\left( \left(\e^{\lambda_{i}^{\mathsf{T}}x}-1\right)
     {\displaystyle\prod\limits_{j=i+1}^{N}}
     \left(1 + \frac{\delta_{j}L_{j-}\left(\e^{\lambda_{i}^{\mathsf{T}}x}-1
\right)}
      {1+\delta_{j}L_{j-}}\right) -\lambda_{i}^{\mathsf{T}}x
\right)F(\cdot,\dx).
\end{align*}
Observe that
\begin{align*}
\prod\limits_{j=1}^{l} \left( 1+w_j \right)
 &= 1 + \sum_{1\leq j\leq l} w_{j}
  + \sum_{1\leq j_{1}<j_{2}\leq l} w_{j_{1}}w_{j_{2}} \\
 &\quad +\sum_{1\leq j_{1}<j_{2}<j_{3}\leq l}w_{j_{1}}w_{j_{2}}w_{j_{3}}
  +...+ \,w_{1}\cdots w_{l}\\
 &= 1 + \sum_{p=1}^{l} S_{p}^{l}(w_{1},...,w_{l}),
\end{align*}
where $S_p^l$ denotes the elementary symmetric polynomial of degree $p$ in $l$
variables, i.e.
\[
S_{p}^{l}(w_{1},...,w_{l})
 := \sum_{1\leq j_{1}<\cdot\cdot\cdot<j_{p}\leq l}
    w_{j_{1}}\cdots w_{j_{p}},\text{ \ \ }1\leq p\leq l.
\]
Thus $\mathbb{B}_i$ may be rearranged as follows:
\begin{align*}
\mathbb{B}_i
 &= \int
\left(\e^{\lambda_{i}^{\mathsf{T}}x}-1-\lambda_{i}^{\mathsf{T}}x\right)F(\cdot,
\dx)
  + \sum_{p=1}^{N-i} \int
   \left( \e^{\lambda_{i}^{\mathsf{T}}x}-1 \right) \,\times\\
 &\qquad\quad
  S_{p}^{N-i}\left( \frac{\delta_{i+1}L_{i+1-}
 \left(\e^{\lambda_{i+1}^{\mathsf{T}}x}-1\right)}{1+\delta_{i+1}L_{i+1-}},\dots,
   \frac{\delta_{N}L_{N-}\left(\e^{\lambda_{N}^{\mathsf{T}}x}-1\right)}
        {1+\delta_{N}L_{N-}}\right) F(\cdot,\dx) \\
 &:= (I)+(II).
\end{align*}
Let us consider in $(II)$ for $p\geq1$ the term
\begin{align*}
&\!\!\int\!\! \left(\e^{\lambda_{i}^{\mathsf{T}}x}-1\right) \!\!
 S_{p}^{N-i} \!\! \left(\!\!
  \frac{\delta_{i+1}L_{i+1-}\left( \e^{\lambda_{i+1}^{\mathsf{T}}x}\!-1\right)}
  {1+\delta_{i+1}L_{i+1-}},\ldots,
  \frac{\delta_{N}L_{N-}\left(\e^{\lambda_{N}^{\mathsf{T}}x}-1\right)}
  {1+\delta_{N}L_{N-}}\!\!\right) F(\cdot,\dx)\\
&= \sum_{i<j_{1}<\cdots<j_{p}\leq N}
 \frac{\delta_{j_{1}}L_{j_{1}-}}{1+\delta_{j_{1}}L_{j_{1}-}}
  \cdots\frac{\delta_{j_{p}}L_{j_{p}-}}{1+\delta_{j_{p}}L_{j_{p}-}} \\
& \qquad\qquad \times\int
    \left(\e^{\lambda_{i}^{\mathsf{T}}x}-1\right)
    \left(\e^{\lambda_{j_{1}}^{\mathsf{T}}x}-1\right)
    \cdots \left(\e^{\lambda_{j_{p}}^{\mathsf{T}}x}-1\right) F(\cdot,\dx).
\end{align*}
With $j_{0}:=i,$ we may write%
\begin{gather}\label{lh}
\left(\e^{\lambda_{i}^{\mathsf{T}}x}-1\right)
\left(\e^{\lambda_{j_{1}}^{\mathsf{T}}x}-1\right)
\cdots\left(\e^{\lambda_{j_{p}}^{\mathsf{T}}x}-1\right) \\
 = (-1)^{p+1}\left(1-\e^{\lambda_{j_{0}}^{\mathsf{T}}x}\right)
             \left(1-\e^{\lambda_{j_{1}}^{\mathsf{T}}x}\right)
       \cdots\left(1-\e^{\lambda_{j_{p}}^{\mathsf{T}}x}\right) \nonumber\\
 = (-1)^{p+1} \left[1+\sum_{q=1}^{p+1}
   S_{q}^{p+1}(-\e^{\lambda_{j_{0}}^{\mathsf{T}}x},\dots,
               -\e^{\lambda_{j_{p}}^{\mathsf{T}}x})\right]
 = (-1)^{p+1} \left[1+(\ast)\right] \nonumber
\end{gather}
where
\begin{align*}
(\ast)
 &= \sum_{q=1}^{p+1} (-1)^{q}
S_{q}^{p+1}(\e^{\lambda_{j_{0}}^{\mathsf{T}}x},\dots,
     \e^{\lambda_{j_{p}}^{\mathsf{T}}x})
  = \sum_{q=1}^{p+1} (-1)^{q} \!\! \sum_{0\leq r_{1}<\cdots<r_{q}\leq p} \!\!
    \e^{\lambda_{j_{r_{1}}}^{\mathsf{T}}x} \cdots
    \e^{\lambda_{j_{r_{q}}}^{\mathsf{T}}x}\\
 &= \sum_{q=1}^{p+1} (-1)^{q} \sum_{0\leq r_{1}<\cdots<r_{q}\leq p}
    \text{ }\underset{O(\left\Vert x\right\Vert^{2})}
     {\underbrace{\left(\e^{\lambda_{j_{r_{1}}}^{\mathsf{T}}x
      +\cdots+\lambda_{j_{r_{q}}}^{\mathsf{T}}x} - 1
 -(\lambda_{j_{r_{1}}}^{\mathsf{T}}x+\cdots+\lambda_{j_{r_{q}}}^{\mathsf{T}}x)
  \right) }}\\
 &\quad+ \sum_{q=1}^{p+1} (-1)^{q} \sum_{0\leq r_{1}<\cdots<r_{q}\leq p}
         \left(1+\lambda_{j_{r_{1}}}^{\mathsf{T}}x+\cdots
               +\lambda_{j_{r_{q}}}^{\mathsf{T}}x\right).
\end{align*}
Obviously, expression \eqref{lh} is of order $O(\left\Vert x\right\Vert^{2})$
for any $p\geq1$, hence (!) it must hold
\[
1+\sum_{q=1}^{p+1} (-1)^{q} \sum_{0\leq r_{1}<\cdots<r_{q}\leq p}
\left(1+\lambda_{j_{r_{1}}}^{\mathsf{T}}x+\cdots+\lambda_{j_{r_{q}}}^{\mathsf{T}
}x\right)
=0.
\]
Therefore, we can deduce the following representation for the integral term
\begin{align}\label{int}
\mathbb{B}_i
&= \int\left(\e^{\lambda_{i}^{\mathsf{T}}x}-1-\lambda_{i}^{\mathsf{T}}x\right)
   F(\cdot,\dx)
 \nonumber\\ &\quad+ \sum_{p=1}^{N-i} \sum_{i<j_{1}<\cdots<j_{p}\leq N}
    \frac{\delta_{j_{1}}L_{j_{1}-}}{1+\delta_{j_{1}}L_{j_{1}-}}
    \cdots\frac{\delta_{j_{p}}L_{j_{p}-}}{1+\delta_{j_{p}}L_{j_{p}-}}
   \sum_{q=1}^{p+1}(-1)^{p+q+1}  \nonumber\\ &\times \!
   \sum_{0\leq r_{1}<\cdots<r_{q}\leq p}  \int \! \left(
  \e^{\left(\lambda_{j_{r_{1}}}+\cdots+\lambda_{j_{r_{q}}}\right)^{\mathsf{T}}x}
   -1-\left(\lambda_{j_{r_{1}}}+\cdots+\lambda_{j_{r_{q}}}\right)^{\mathsf{T}}\!
   x\right) F(\cdot,\dx)\nonumber \nonumber\\
&= \widehat\kappa(\lambda_i)
+ \sum_{p=1}^{N-i} \sum_{i<j_{1}<\cdots<j_{p}\leq N}
    \frac{\delta_{j_{1}}L_{j_{1}-}}{1+\delta_{j_{1}}L_{j_{1}-}}
    \cdots\frac{\delta_{j_{p}}L_{j_{p}-}}{1+\delta_{j_{p}}L_{j_{p}-}}
\nonumber\\   &\qquad\qquad\quad \times\sum_{q=1}^{p+1}(-1)^{p+q+1}
   \sum_{0\leq r_{1}<\cdots<r_{q}\leq p} \widehat\kappa
    \left(\lambda_{j_{r_{1}}}+\cdots+\lambda_{j_{r_{q}}}\right).
\end{align}

\subsection{First order expansion of \eqref{int}}
Let us consider the first order expansion of $\mathbb{B}_i$; we get
\begin{align*}
\mathbb{B}_i
 &= \widehat\kappa(\lambda_i)
 + \sum_{i<j<n} \frac{\delta_{j}L_{j-}}{1+\delta_{j}L_{j-}}
   \sum_{q=1}^{2} (-1)^{q} \sum_{0\leq r_{1}<\cdots<r_{q}\leq1}
    \widehat\kappa \left(\lambda_{j_{r_{1}}}+\cdots+\lambda_{j_{r_{q}}}\right)
 \\ &\quad+ O(\left\Vert L\right\Vert^{2}).
\end{align*}
Note that
\begin{align*}
& \sum_{q=1}^{2} (-1)^{q} \sum_{0\leq r_{1}<\cdots<r_{q}\leq1}
  \widehat\kappa\left(\lambda_{j_{r_{1}}}+\cdots+\lambda_{j_{r_{q}}}\right) \\
&\quad = - \sum_{0\leq r_1\leq1}\text{}\widehat\kappa
            \left(\lambda_{j_{r_1}}\right)
 \,\,+ \sum_{0\leq r_{1}<r_{2}\leq1}
    \widehat\kappa\left(\lambda_{j_{r_{1}}}+\lambda_{j_{r_{2}}}\right) \\
&\quad = - \widehat\kappa\left(\lambda_{j_0}\right)
    - \widehat\kappa\left(\lambda_{j_1}\right)
    + \widehat\kappa\left(\lambda_{j_0}+\lambda_{j_1}\right).
\end{align*}
Thus we obtain the following expression for the first order expansion of the
integral term $\mathbb{B}_i$
\begin{align}\label{fo_int}
\mathbb{B}_i'
&= \widehat\kappa(\lambda_i)
 + \sum_{i<j\leq N}\frac{\delta_{j}L_{j-}}{1+\delta_{j}L_{j-}}
  \big(\widehat\kappa(\lambda_i+\lambda_j)-\widehat\kappa(\lambda_i)
       -\widehat\kappa(\lambda_j)\big),
\end{align}
which leads to the following approximation for the drift term $b_i$ in
\eqref{dr}
\begin{align}\label{fo_int_all}
b_i'
&= \kappa(\lambda_i)
 + \sum_{i<j\leq N}\frac{\delta_{j}L_{j-}}{1+\delta_{j}L_{j-}}
  \big(\kappa(\lambda_i+\lambda_j)-\kappa(\lambda_i)
       -\kappa(\lambda_j)\big),
\end{align}
taking also the terms stemming from the diffusion into account.

\subsection{Second order expansion of \eqref{int}}
Analogously, we can also derive a second order expansion of $\mathbb{B}_i$; we
get
\begin{align*}
\mathbb{B}_i
 &= \widehat\kappa(\lambda_i)
  + \sum_{i<j\leq N}\frac{\delta_{j}L_{j-}}{1+\delta_{j}L_{j-}}
  \big(\widehat\kappa(\lambda_i+\lambda_j)-\widehat\kappa(\lambda_i)
       -\widehat\kappa(\lambda_j)\big)
 \\ &\quad+ \sum_{i+1\le k<l\le N}
         \frac{\delta_k L_{k-}}{1+\delta_k L_{k-}}
         \frac{\delta_l L_{l-}}{1+\delta_l L_{l-}}
  \Big(\widehat\kappa(\lambda_i+\lambda_k+\lambda_l)
    - \widehat\kappa(\lambda_i+\lambda_k) \notag
 \\ &\qquad\qquad\qquad - \widehat\kappa(\lambda_i+\lambda_l)
    - \widehat\kappa(\lambda_k+\lambda_l)
    + \widehat\kappa(\lambda_i) + \widehat\kappa(\lambda_k)
    + \widehat\kappa(\lambda_l)\Big)
 \\ &\quad+ O(\left\Vert L\right\Vert^{3}),
\end{align*}
which leads to the following second order expansion of $b_i$ in \eqref{dr}
\begin{align}\label{so_int}
b_i''
&= \kappa(\lambda_i)
 + \sum_{i<j\leq N}\frac{\delta_{j}L_{j-}}{1+\delta_{j}L_{j-}}
  \big(\kappa(\lambda_i+\lambda_j)-\kappa(\lambda_i)
       -\kappa(\lambda_j)\big) \nonumber
 \\ &\quad+ \sum_{i+1\le k<l\le N}
         \frac{\delta_k L_{k-}}{1+\delta_k L_{k-}}
         \frac{\delta_l L_{l-}}{1+\delta_l L_{l-}}
  \Big(\widehat\kappa(\lambda_i+\lambda_k+\lambda_l)
    - \widehat\kappa(\lambda_i+\lambda_k) \notag
 \\ &\qquad\qquad - \widehat\kappa(\lambda_i+\lambda_l)
    - \widehat\kappa(\lambda_k+\lambda_l)
    + \widehat\kappa(\lambda_i) + \widehat\kappa(\lambda_k)
    + \widehat\kappa(\lambda_l)\Big).
\end{align}

\section{Derivation of \eqref{dZ}}
\label{DerZ}

Using the It\^o formula for general semimartingales (cf. \cite[Theorem
I.4.57]{JacodShiryaev03}) we have
\begin{align}\label{sum}
Z_{j}
 &= Z_{j}(0) + \int_{0}^{\cdot}f^{\prime}(G_{j}(s-)) \ud G_{j}
  + \frac{1}{2}\int_{0}^{\cdot}f^{\prime\prime}(G_{j})
     \ud \langle G_{j}^c,G_{j}^{c}\rangle  \nonumber\\
 &\quad+ \sum_{0<s\leq\cdot} \left(f(G_{j}(s))-f\left(G_{j}(s-)\right)
      -f^{\prime}\left(G_{j}(s-)\right)  \Delta G_{j}(s)\right),
\end{align}
where $\langle G_{j}^c,G_{j}^{c}\rangle$ denotes the quadratic variation of the
continuous martingale part of $G_{j},$ that is
\begin{equation}\label{cv}
\ud\langle G_{j}^c,G_{j}^{c}\rangle(s)
 = \left\vert \lambda_{j} \right\vert^{2}(s)\alpha(s) \ds.
\end{equation}
The sum in \eqref{sum}, using \eqref{G-triplet}, may be written as
\begin{align}\label{js}
& \sum_{0<s\leq\cdot}
 \left( f(G_{j}(s-)+\Delta G_{j}(s))
   - f\left(G_{j}(s-)\right)
   - f^{\prime}\left(  G_{j}(s-)\right) \Delta G_{j}(s)\right)\\
&= \int_{0}^{\cdot}\int_{\R^m}
   \left( f(G_{j}(s-)+\lambda_{j}^{\mathsf{T}}x)
     - f\left( G_{j}(s-) \right)
     - f^{\prime}\left( G_{j}(s-) \right)\lambda_{j}^{\mathsf{T}}x\right)
   \mu\dsdx\nonumber\\
&=\int_{0}^{\cdot}\int_{\R^m}
  \left( f(G_{j}(s-)+\lambda_{j}^{\mathsf{T}}x)
   - f\left( G_{j}(s-) \right)
   - f^{\prime}\left( G_{j}(s-) \right) \lambda_{j}^{\mathsf{T}}x\right)
   F(s,\dx)\ds\nonumber\\
&\,\,\, + \int_{0}^{\cdot}\int_{\R^m}
  \left( f(G_{j}(s-)+\lambda_{j}^{\mathsf{T}}x)
   - f\left( G_{j}(s-) \right)
   - f^{\prime}\left( G_{j}(s-) \right)\lambda_{j}^{\mathsf{T}}x\right)
   \mt\dsdx.  \notag
\end{align}
Moreover,
\begin{align}\label{fp}
\int_{0}^{\cdot}f^{\prime}(G_{j}(s-))\ud G_{j}
 &= \int_{0}^{\cdot} f^{\prime}(G_{j}(s-))b_{j}\ds
  + \int_{0}^{t}f^{\prime}(G_{j}(s-))\sqrt{\alpha}\lambda_{j}^{\mathsf{T}}
     \ud W\notag\\
 &\quad+ \int_{0}^{\cdot}\int_{\R^m} f^{\prime}(G_{j}(s-))
     \lambda_{j}^{\mathsf{T}}x \,\mt\dsdx.
\end{align}
Finally, by plugging \eqref{cv}, \eqref{js} and \eqref{fp} into \eqref{sum},
\eqref{dZ} follows.

\section{Derivation of \eqref{SDEY}}
\label{DerY}

The computations are completely analogous to Appendix \ref{DerZ}, thus omitted
for brevity. The
coefficients of $Y_\kl$ in \eqref{SDEY} are
\begin{align}
A_\kl(s,L(s))
 &= \sum_{i=k,l} g_i(G_{k}(s-),G_l(s-)) b_i(s) \nonumber\\
 &\qquad + \frac12 \sum_{i,j=k,l} \int_{0}^{\cdot} g_{ij}(G_{k}(s-),G_l(s-))
     \lambda_i(s) \lambda_{j}(s) \alpha(s) \nonumber\\
 & + \int_{\R^m}
    \bigg(  g(G_{k}(s-)+\lambda_k^{\mathsf{T}} x,G_l(s-)+\lambda_l^{\mathsf{T}}
x)
          - g(G_{k}(s),G_{l}(s)) \nonumber\\
 &\qquad\qquad -  \sum_{i=k,l} g_i(G_{k}(s-),G_l(s-)) \lambda_i^{\mathsf{T}} x
\bigg)
       F(s,\dx), \\
B_\kl^{\mathsf{T}}(s,L_\kl(s))
 &= \sum_{i=k,l} g_i(G_{k}(s-),G_l(s-))
\sqrt{\alpha(s)}\lambda_i^{\mathsf{T}}(s)\\
\intertext{and}
C_\kl(s,L_\kl(s),x)
 &= \sum_{i=k,l} g_i(G_{k}(s-),G_l(s-)) \lambda_i^{\mathsf{T}}(s)x \nonumber\\
 &\quad + g\big(G_{k}(s-)+\lambda_k^{\mathsf{T}}
x,G_l(s-)+\lambda_l^{\mathsf{T}} x\big)
    - g\big(G_{k}(s),G_{l}(s)\big) \nonumber\\
 &\qquad -  \sum_{i=k,l} g_i(G_{k}(s-),G_l(s-)) \lambda_i^{\mathsf{T}} x.
\end{align}
Here $L_{kl}(s):=(L_k(s),L_l(s))$ and denotes that $B_\kl$ and $C_\kl$ depend on
$L_k$ and $L_l$ (via $G_k$ and $G_l$).

\bibliographystyle{alpha}
\bibliography{references}

\end{document}